\documentclass[preprint]{elsarticle}
\usepackage{graphicx,bm,epsf,float,amssymb}
\usepackage[font={scriptsize}]{subcaption} 

\usepackage{lineno}
\usepackage{multirow}

\usepackage{amsmath}
\usepackage{xspace}
\newcommand{\um}[1]{\ensuremath{#1~\mu\mathrm{m}}\xspace}

\begin{document}

\begin{frontmatter}
\title{Interaction position, time, and energy resolution in organic scintillator bars with dual-ended readout}
\author[sandia-ca]{M. Sweany\corref{cor1}}
\author[hawaii]{A. Galindo-Tellez}
\author[sandia-ca]{J. Brown}
\author[sandia-ca]{E. Brubaker}
\author[hawaii]{R. Dorrill}
\author[hawaii]{A. Druetzler}
\author[hawaii]{N. Kaneshige}
\author[hawaii]{J. Learned}
\author[hawaii]{K. Nishimura}
\author[hawaii]{W. Bae}

\address[sandia-ca]{Sandia National Laboratory, Livermore, CA 94550, USA}
\address[hawaii]{University of Hawai`i at M\={a}noa, Honolulu, HI 96822, USA }

\cortext[cor1]{Corresponding author: msweany@sandia.gov}

\begin{abstract}
We report on the position, timing, and energy resolution of a range of plastic scintillator bars and reflector treatments using 
dual-ended silicon photomultiplier readout. These measurements are motivated by the upcoming construction of an 
optically segmented single-volume neutron scatter camera, in which neutron elastic scattering off of hydrogen is used to
kinematically reconstruct the source direction and energy of an incoming neutron. For this application, interaction position resolutions 
of about 10~mm and timing resolutions of about 1~ns are necessary to achieve the desired efficiency for fission-energy neutrons. 
The results presented here indicate that this is achievable with an array of $5\times5\times190~\mathrm{mm}^3$ bars of EJ-204 scintillator wrapped in Teflon tape,
read out with SensL's J-series $6\times6~\mathrm{mm}^2$ silicon photomultipliers. With two independent setups, we also explore the systematic 
variability of the position resolution, and show that, in general, using the difference in the pulse arrival time at the two ends is less susceptible to systematic variation
than using the log ratio of the charge amplitude of the two ends. Finally, we measure a bias in the absolute time of interactions
as a function of position along the bar: the measured interaction time for events at the center of the bar is $\sim$100 ps later than interactions near the SiPM.

\end{abstract}

\begin{keyword}
fast neutron imaging \sep special nuclear material detection
\end{keyword}

\end{frontmatter}

\section{Introduction}
\label{sec:one}
Recently, the concept of a single volume neutron scatter camera was reported~\cite{pips, svsc, mtc}. 
In comparison to current implementations of the technology (see e.g.~\cite{miner}), a single volume scatter 
 camera aims to detect multiple neutron-proton interactions within the same contiguous volume of 
 scintillator, rather than distributed volumes of scintillator. If successful, the resulting instrument would 
 out-perform current state-of-the-art double-scatter imagers in terms of imaging efficiency for fission-energy neutrons by an 
 estimated order of magnitude~\cite{svsc}, as well as improve deployability factors such as size, weight, 
 and power consumption. For some non-proliferation applications, the reduced size of the instrument would also enable closer inspection of objects, leading 
 to rate increases and corresponding reductions in acquisition times and an improvement in imaging 
 resolution. 

The key to realizing such an instrument lies in accurately reconstructing the time, position, and energy 
of neutron-proton interactions within the volume of scintillator. The kinetic reconstruction of the neutron 
direction requires two neutron-proton elastic scatters: the neutron direction before the first scatter is 
constrained to a cone defined by the angle $\theta$:
\begin{linenomath*}
\begin{equation}
\mathrm{cos}(\theta) = \sqrt{\frac{E_{n'}}{E_n}},
\end{equation}
\end{linenomath*}
where $E_n$ is the incoming neutron energy and $E_{n'}$ is the neutron's energy after the first 
interaction. In the non-relativistic limit, $E_{n'}$ is determined by the neutron time-of-flight ($\Delta t$) and distance ($\Delta d$)
between the first and second neutron interactions, where $m_n$ is the mass of the neutron:
\begin{linenomath*}
\begin{equation}
E_{n'} = \frac{1}{2} m_n \left(\frac{\Delta d}{\Delta t} \right)^2.
\end{equation}
\end{linenomath*}
Finally, $E_n$ is determined by the energy deposited in the first interaction, $E_p$, measured by the 
light emitted in the first scintillation pulse, and $E_{n'}$:
\begin{linenomath*}
\begin{equation}
E_n = E_{n'} + E_p.
\end{equation}
\end{linenomath*}
The accuracy of the cone defined by $\theta$ is thus dependent on the energy resolution of the first 
scintillation pulse, and the timing and position resolution of the two interactions. The focus of this work is to characterize with 
experimental data the energy, timing, and position resolution of an optically segmented volume of 
scintillator.  In this instance of the concept, only the position along each bar segment of the array must 
be reconstructed for the position of each interaction, while the other position dimensions are determined by the 
particular photodetector in which the scintillation light was detected. Monte Carlo simulations of various 
detector configurations of optically segmented bars have been previously performed~\cite{pips} and 
provided guidance as to the optimal choice of scintillator, reflector material, and photon detection 
technology. However, practical considerations have limited the experimental space explored. 
  
There are other detection systems that share the same goals to optimize the energy, timing, and 
position resolution of interactions within scintillator volumes. Time-of-flight Positron Emission Tomography (TOFPET)
systems require good spatial and timing resolution, and benefit from increased light output from gamma 
interactions in the scintillator volume  (see e.g.~\cite{tofpet}): because only gamma interactions are of interest, inorganic 
scintillators are used, which typically have higher light output compared to organic scintillators. While 
TOFPET systems can look very similar to what is described here, the problem of reconstructing a second 
interaction very close in time and position to the first motivates scintillator responses not only with fast 
pulse rise times, but with reasonably fast pulse decay times as well. PET systems also work in a relatively limited 
energy range, annihilation gammas at 511 keV, and therefore use the energy 
measurement primarily for background rejection as opposed to reconstruction. Neutrino detection 
systems also benefit from good energy resolution to reconstruct the incoming neutrino spectrum, and good spatial and 
timing resolution, either to reconstruct the neutrino direction (for neutrino-electron scattering 
interactions) or for background rejection based on event topology (for inverse-beta decay interactions). 
Because neutrino interaction cross sections are so low, large detector systems are a paramount design 
consideration, motivating larger segmentation and cost-effective readout. These requirements often lead to similar
systems with dual-sided readout of long canes with a similar aspect ratio presented here (see e.g.~\cite{sweany, panda, prospect}). 
While these differences may result in different design details, the methods and conclusions presented here are relevant to 
a wider detection community.

\subsection{Design Considerations}
The experimental measurements presented here are constrained to 190 mm long rectangular scintillator 
bars with a cross sectional area of $5\times5~\mathrm{mm}^2$. The length constraint is determined 
by the approximate desired volume of the 
instrument, coupled with the currently available readout size of fast photo-detectors (approximately 50 
mm). For photo-detection, the fastest available on the market currently are micro-channel plate 
photomultiplier tubes (MCP-PMTs) and silicon photomultipliers (SiPMs).  In terms of timing, MCP-PMTs typically have 
the best single photoelectron timing performance, but have a higher cost and typically much lower photodetection efficiency (PDE). 
SiPMs were chosen for their low cost and superior photon separation and PDE.
For data acquisition, we use the 4-channel DRS4 evaluation board~\cite{drs41, drs42} from the Paul Scherrer Institut (PSI) for its high sampling speed 
and ease of use. 

The scintillator types explored here are constrained to the plastic scintillators with parameters 
from the manufacturer that are expected to optimize our performance metrics. In this case, the relevant 
scintillator parameters for consideration are: 
\begin{enumerate}
\item total output and emission wavelength well-matched to the wavelength-dependent PDE of the 
SiPMs to optimize detectable photons,
\item rise time of the scintillator pulse for optimal timing resolution as well as position resolution along 
the bar,
\item emission wavelength in relation to surface treatment reflectivity, and
\item attenuation length ($\lambda$) to either optimize position resolution (short) or maximize the 
number of detectable photons (long).
\end{enumerate}

The method used to reconstruct the position resolution dictates the desired attenuation length. The 
methods used here are based on the difference in the pulse arrival time and amplitude at each end of the bar. 
Potentially, the difference in pulse shape provides an additional method~\cite{pips}, however this is not explored in this work. 
In the case of the pulse 
amplitude difference, a shorter attenuation length maximizes the difference in amplitude and therefore the 
position resolution. However, this leads to a reduction in energy resolution. The difference in both pulse arrival time and pulse shape may be enhanced by a longer 
attenuation length: in the case of pulse arrival time, an increase in photons arriving directly from the 
interaction could improve the precision of the measurement, and in the case of the pulse shape, an 
increase in late arriving photons may enhance the tail end of the pulse for interactions further from the 
readout end. Based on this, the preference is for longer attenuation lengths. It should be noted, however, 
that different attenuation lengths for different processes may exist within a bar, and so measurements
may need to be re-evaluated if the bar length changes significantly.

\begin{table}\begin{center}
\begin{tabular}{|c||c|c|c|c|}
\hline
Scintillator	&$t_R$ (ns)	&$\lambda$ (cm$^{-1}$)	&$N_e$ (MeV$^{-1}$)	&$N_D$ (MeV$^{-1}$) \\	
\hline
{\bf EJ-200}&0.9			&380					&10,000				&4,905						\\
{\bf EJ-204}&0.7		&160					&10,400				&5,084						\\
EJ-208	&1.0			&400					& 9,200				&4,378						\\
{\bf EJ-230}&0.5			&120					&9,700				&4,557						\\
EJ-232	&0.35		&-					&8,400				&3,679						\\
EJ-260	&-			&350					&9,200				&3,470						\\
EJ-262	&-			&250					&8,700				&3,548						\\
{\bf EJ-276}&-			&-					&8,600				&4,203						\\
EJ-276G	&-			&-					&8,000				&2,991						\\
\hline
\end{tabular}
\caption{A list of plastic scintillator candidates from Eljen Technology, with the parameters listed from their website. Some parameters were not listed. In addition, the number of detectable photons per MeV after a PDE is applied, $N_D$,  is listed for $6\times6~\mathrm{mm}^{2}$ J-series SiPM from SensL. Bold entries correspond to those with the highest $N_D$ values.}
\label{tab:tab1}
\end{center}
\end{table}

Table \ref{tab:tab1} lists the available plastic scintillators from Eljen Technology's catalog that have been considered 
for our bar length. Not included in the list are EJ-212 and EJ-214, which are intended for thin sheets, and  
EJ-228, which has an attenuation length too small for bars greater than 10 cm in length. 
Loaded scintillators such as EJ-254 (boron) and EJ-256 (lead) are also not included, nor are the high-temperature 
analogs of EJ-200 and EJ-208 (EJ-244/M and EJ-248/M). The number of detectable photons, 
$N_D$, for SensL's J-series SiPMs was determined by sampling the total emitted photons, $N_e$, over 
the emission wavelength of the scintillator, and applying the wavelength-dependent PDE for a 5.0 V 
over-voltage. No propagation losses or geometrical acceptance effects are included in this estimate of 
$N_D$. The top three candidates in terms of $N_D$ are EJ-200, EJ-204, and EJ-230, shown in bold in 
Table \ref{tab:tab1}. We are also including EJ-276, for its potentially beneficial neutron/gamma 
pulse-shape discrimination power: neutrons and gammas can also be discriminated by the time-of-flight 
between interactions, but the power of that metric will be dependent on the timing resolution of the final 
system.  

In terms of surface reflectivity, we have narrowed the options to 3M's Enhanced Specular Reflector 
(ESR), which has a specular reflectivity of 98.5\% above approximately 380 nm wavelengths 
\cite{janecek, motta}, PTFE (Teflon) tape with a diffuse reflectivity greater than 95\% over a broad 
wavelength range for a 0.5 mm thick layering \cite{janecek, teflon}, and finally no surface treatment with only total internal reflection (TIR) to guide the 
light to the readout ends. These were chosen because they are among the highest 
reflectivity materials for pure diffuse (Teflon) and specular (ESR) reflectivity that are commercially 
and readily available.

In addition to surface reflectivity, the surface roughness of the scintillator material has been previously 
shown by others to have a significant effect on the light transportation properties. On results 
with a $6\times6\times200~\mathrm{mm}^{3}$ bar of BC400, the authors of~\cite{gierlik} report greater 
than an order of magnitude reduction in the number of photons emitted at 200 mm from the 
photodetector surface compared to 20 mm from the photodetector surface for a bare bar with poor 
surface roughness: this value is reduced to 30\% for a finely polished sample. The effect is also evident 
with highly reflective surface coatings: the ratio of detected photoelectrons between 200 and 20 mm 
was reported to be 60\% for poorly polished samples wrapped with ESR, and only 9\% for a finely 
polished sample. While there was some effort to reproduce these effects here, polishing is a time 
consuming processes, especially for bars of this form factor. For the studies reported here, we used the best surface polish quality available from the scintillator supplier.

\section{Experimental Setup}
\label{s:detchar}

Two independent experimental setups are used to perform characterizations on the test scintillators.  
These two setups serve to validate one another, and also provide an effective method to identify and 
study systematic effects that may lead to differences in performance.

\begin{figure}[!htbp]
	\begin{subfigure}[h]{0.45\columnwidth} 
		\centering
		\includegraphics[width=\columnwidth]{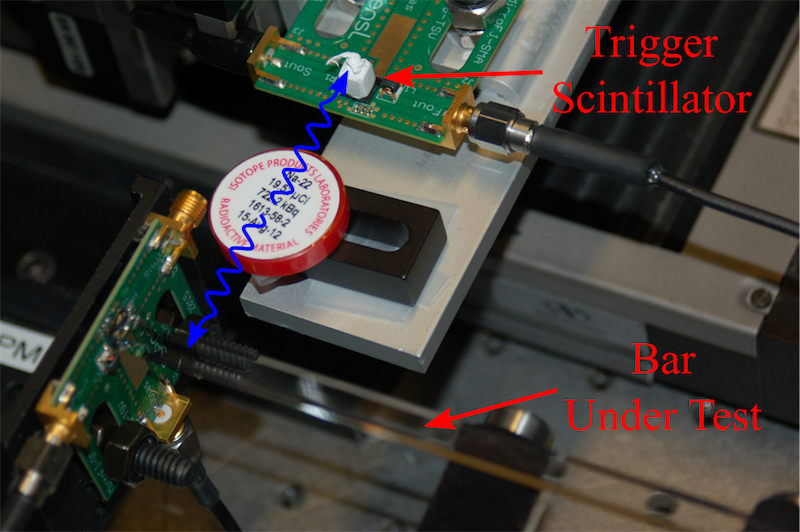}
		\caption{}
	\end{subfigure}	
	\begin{subfigure}[h]{0.53\columnwidth} 
		\centering
		\includegraphics[width=\columnwidth]{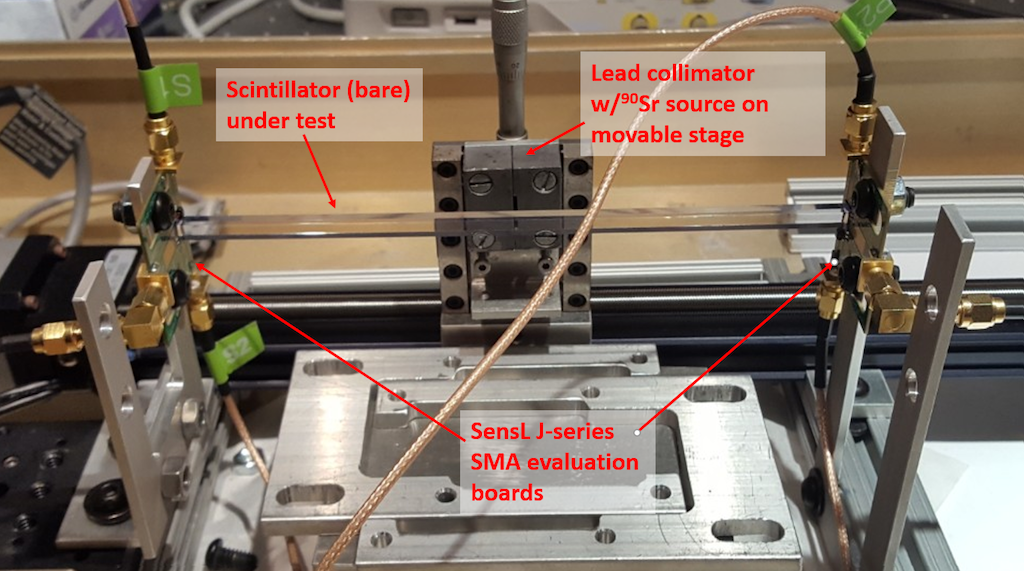}
		\caption{}
	\end{subfigure}
\caption{A picture of the testing apparatus at (a) SNL and (b) UH. For (a), the blue lines indicate coincident back-to-back 511 gammas from the $^{22}$Na source interacting
in both the bar under test and the trigger scintillator.}
\label{fig:fig1} 
\end{figure}

At Sandia National Laboratories (SNL), timing, position resolution, and energy resolution are all characterized using 
the back-to-back 511~keV annihilation gammas 
emitted from a $^{22}$Na source. A $5 \times 5 \times 5~\mathrm{mm}^{3}$ stilbene crystal wrapped in 
Teflon and coupled to a single-channel SensL J-series $6\times6~\mathrm{mm}^{2}$ SiPM is used to 
trigger the data acquisition system. The SiPM is biased at +30~V using a BK Precision 1761 DC power 
supply. This ``trigger'' scintillator is placed on two automated linear stages from Newport: UTS150PP, 
with a 150~mm range and UTS50PP, with a 50~mm range. Both stages are controlled through a GPIB 
interface with Newport's MM3000 motion controller. The manufacturer quotes a guaranteed \um{4} 
accuracy with a \um{0.5} repeatability for both linear stages, far exceeding the resolution limit of our 
apparatus: 5/$\sqrt{12}$~mm, due to the size of the trigger scintillator. The 150~mm stage is used to scan the length 
of the bar, and the 50~mm stage is used for vertical alignment.  The $5 \times 5 \times 190$~mm ``test'' scintillator bar is placed 
opposite the trigger scintillator, with the long axis of the test bar parallel to the motion of the 150~mm 
stage. The test scintillator bar is coupled with optical grease to two additional single-channel SensL J-series 
$6\times6~\mathrm{mm}^2$ SiPMs, also biased at +30 V. Two cylindrical optical rods wrapped with black felt
are used to support the scintillator bar in order to prevent decoupling from the SiPMs. Finally, the $^{22}$Na source is placed 
equidistant between the trigger and test scintillators and on the same stage as the trigger scintillator, so 
that each measurement has the same distribution of interactions in the test bar, and there is a one-to-one 
translation between the trigger scintillator position and the interaction location in the test scintillator. 
The entire apparatus, shown in Figure \ref{fig:fig1}a, is placed in a light-tight enclosure.  The energy resolution
is characterized at the center of the bar, and the 
position and timing resolution are characterized with a set of trigger scintillator positions that 
span the bar from 15 mm from one edge of the test bar to 135~mm from the edge. For two scans (EJ-200 bare1 and EJ-204 ESR), 
the total range covers over half the bar, but the spacing and range differs from the others. For each position, 100,000 
triggers are acquired with the DRS4 Eval board, which is calibrated for 5.12~GHz sampling.  We acquire full 
waveforms, and use C++ along with libraries from the analysis toolkit ROOT~\cite{root} for pulse 
processing and subsequent analysis. 

At the University of Hawaii (UH), position resolutions are studied using a $^{90}$Sr (beta-emitting) source. 
A lead collimator is used to 
limit the incident position of betas from the source onto a narrow region of the scintillator. The
collimator geometry limits the range of incident positions of betas on the scintillator to under 2~mm. 
The entire collimator is mounted to a Velmex XSlide (XN10-0120-M01-71), driven by a stepper motor (PK245-01AA), so that 
the position of incidence can be scanned over the length of the scintillator bar under test. Source 
positions are varied by 5~mm per step, via a USB interface to the motor controller. The 
manufacturer-specified repeatability and step sizes are well below our resolution, as well as the geometric spread 
induced by the collimator. The photodetectors, including bias voltage, and data acquisition electronics 
are identical to the SNL setup, though the waveform analysis tools to process the stored data were 
independently developed at each site.  

For each source position, a total of 4000 interactions are acquired. The 
DRS4 evaluation board triggers itself on a coincident signal on both SiPMs, with a trigger threshold of 
roughly 7~mV. The UH setup is shown in Figure~\ref{fig:fig1}b. At UH, energy calibrations and
resolution measurements are obtained by placing a separate $^{137}$Cs source at the center of the bar.

\section{Waveform Analysis}
\begin{figure}[!htbp]
	\begin{subfigure}[h]{\columnwidth} 
		\centering
		\includegraphics[width=\columnwidth]{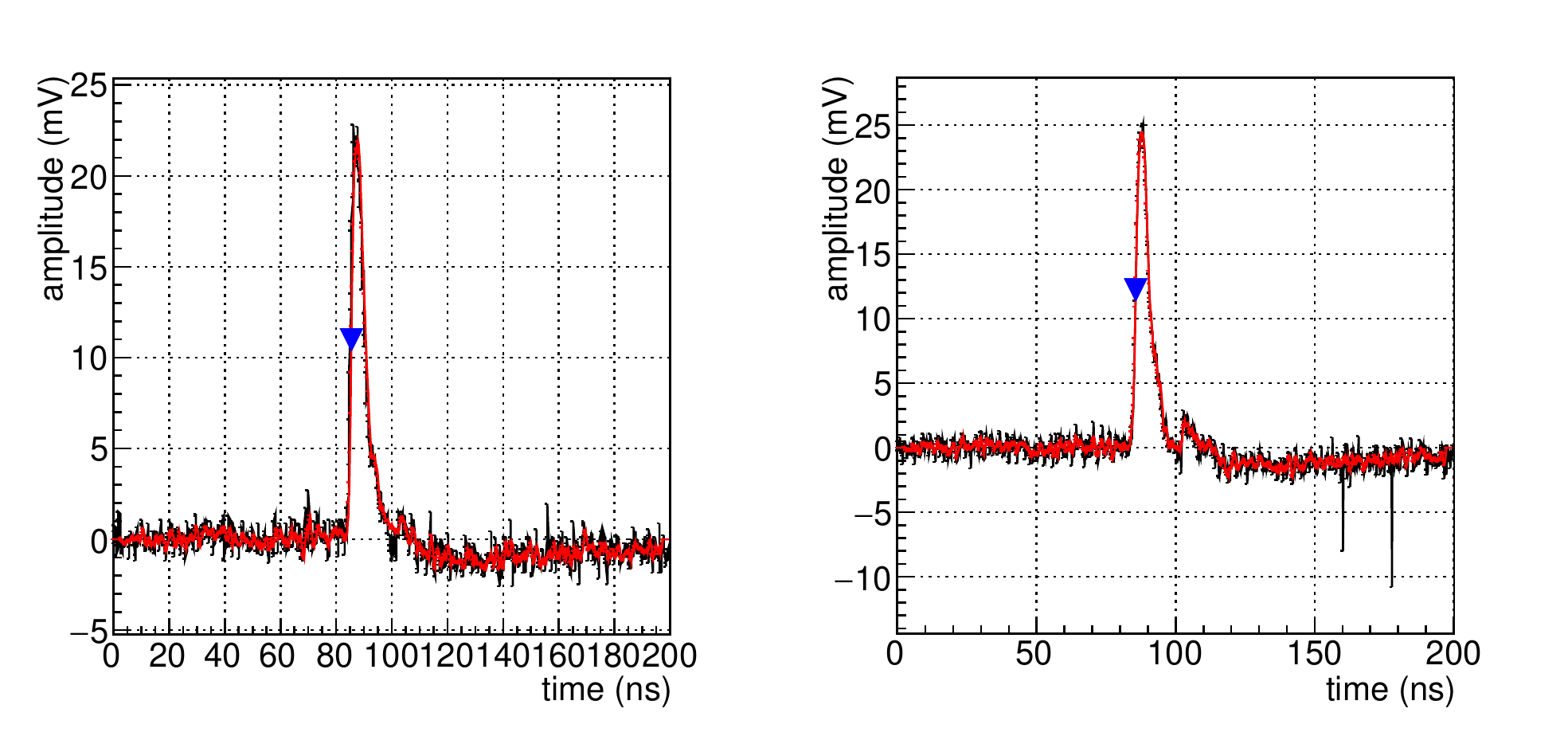}
	\end{subfigure}	
	\begin{subfigure}[h]{\columnwidth} 
		\centering
		\includegraphics[width=\columnwidth]{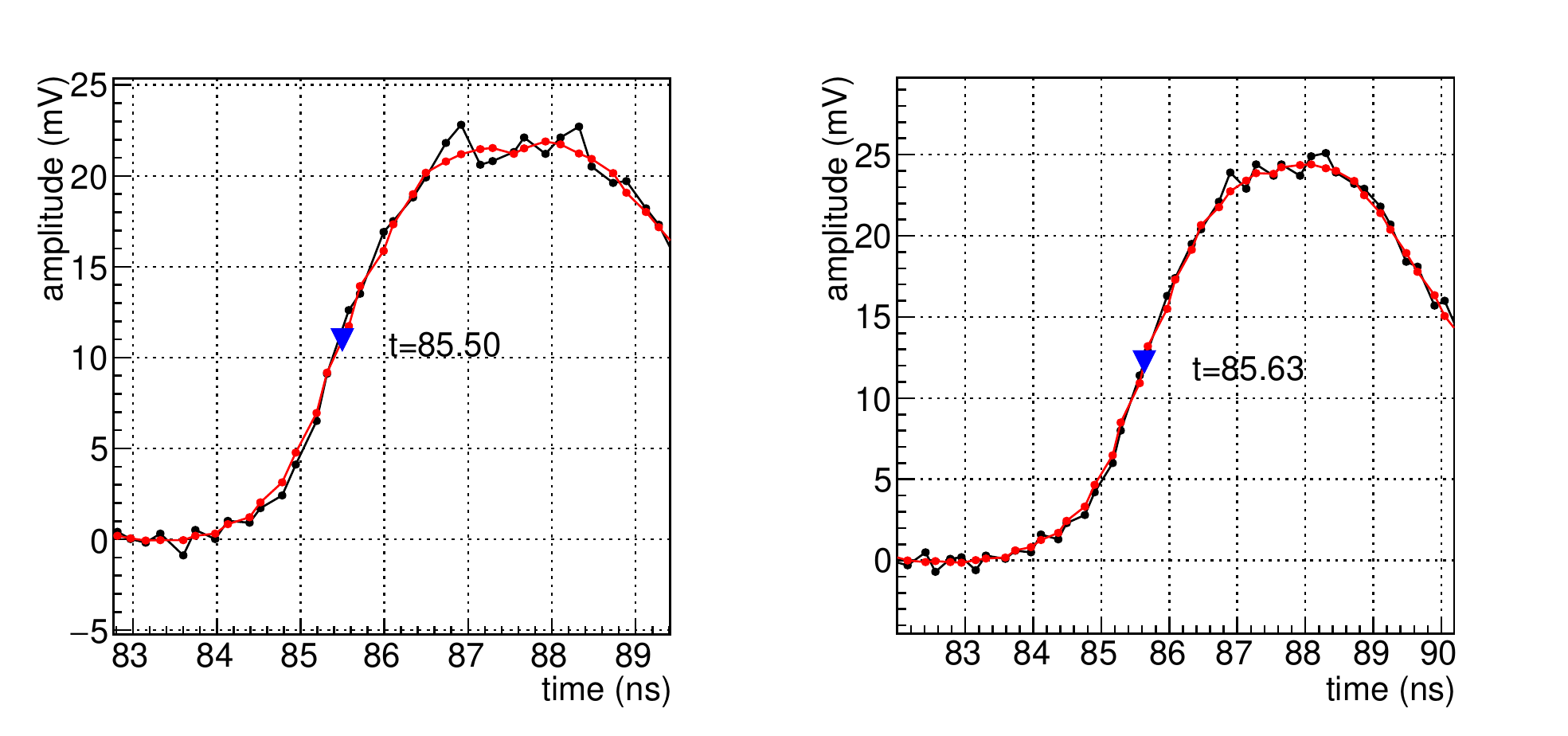}
	\end{subfigure}
\caption{(top) An example waveform from the bar channels coupled to EJ-204 with no reflector material. 
The black curve is the waveform output from the DRS4, and the red curve is after spike-removal and 
smoothing filters are applied. The pulse time is shown with the blue triangle. (bottom) The same two 
traces zoomed into the region of 50\% of the pulse rise time, with sample values shown with dots and 
the pulse time indicated on the figure. The spikes between 150 and 200 ns for channel two are removed with
the filtering.}
\label{fig:fig2} 
\end{figure}

For all measurements, we acquire the fast output (FOUT) of the SiPMs, primarily due to the finite number of samples that the DRS4 can record: 
for 5 GS/s digitization, trace lengths are limited to 200 ns. 
Waveforms recorded with the DRS4 are analyzed to determine the pulse height and time. 
Figure \ref{fig:fig2} shows two example waveforms from the two readout SiPMs of a scintillator bar, in 
this case EJ-204 with no reflector material. 
The pulse shape from the FOUT of the SiPM typically has a 
significant dip below baseline, shown in the figures around 100 ns: for this reason we take the pulse height, 
or maximum sample value after baseline subtraction, for the energy measurement. In addition, the DRS4 output often includes one or 
two sample spikes that are removed by considering the mean before and after a particular 
sample value. Finally, a smoothing filter is applied. 

The pulse time is the interpolated value between samples corresponding to 50\% of the maximum value on 
the rising edge of the filtered pulse. This timing algorithm was found to outperform methods based on the zero crossing of the derivative. 
Figure \ref{fig:fig2} shows the original waveform output from the 
DRS4, the result after spike removal and smoothing is applied, and the pulse time. The same trace zoomed 
into the pulse time is also shown with the samples indicated by dots on the filtered output. 

\section{Detector Response}
\subsection{Energy response}
\begin{figure}[!htbp]
	\begin{subfigure}[h]{0.49\columnwidth} 
		\centering
		\includegraphics[width=\columnwidth]{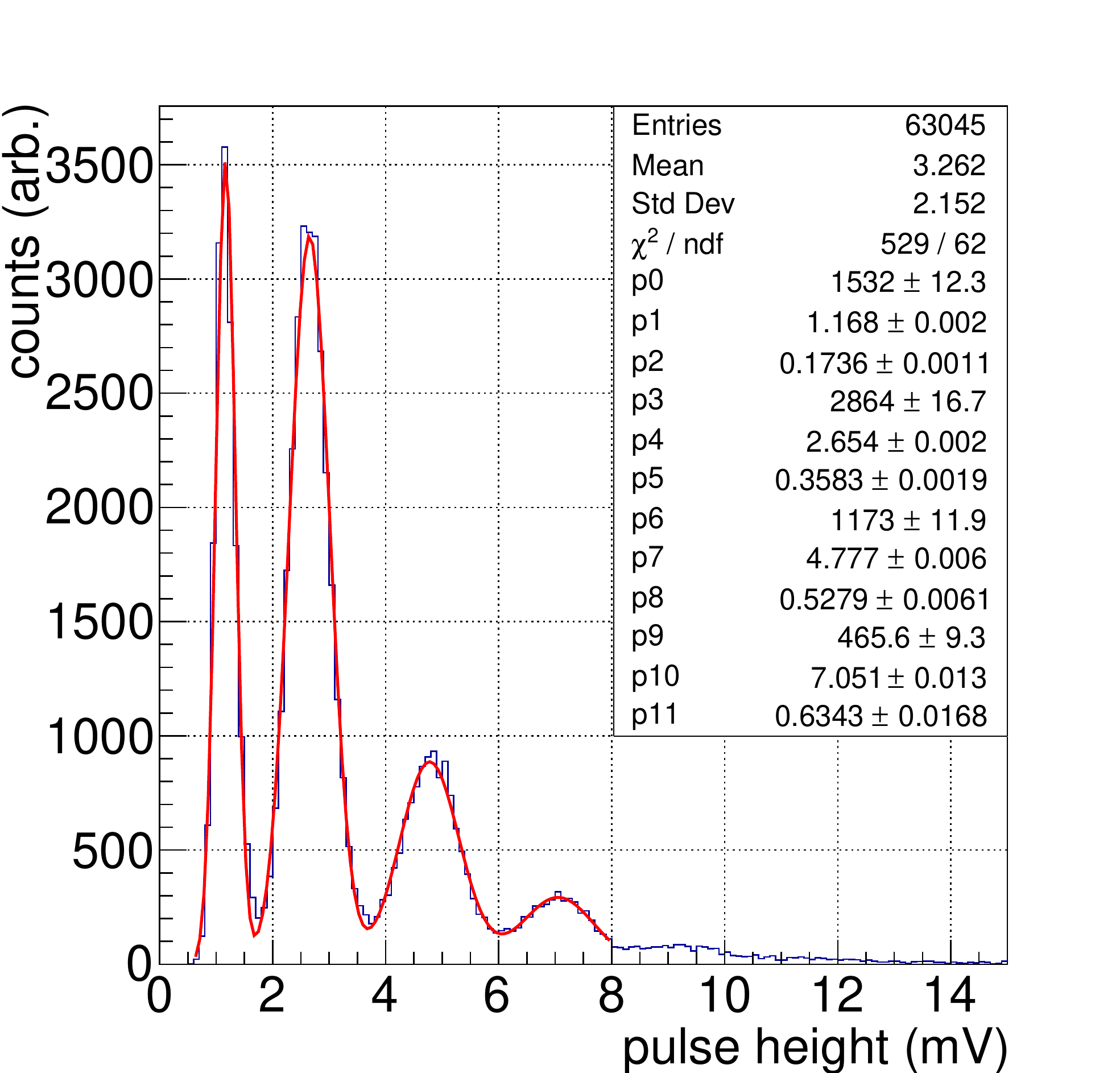}
		\caption{}
	\end{subfigure}	
	\begin{subfigure}[h]{0.49\columnwidth} 
		\centering
		\includegraphics[width=\columnwidth]{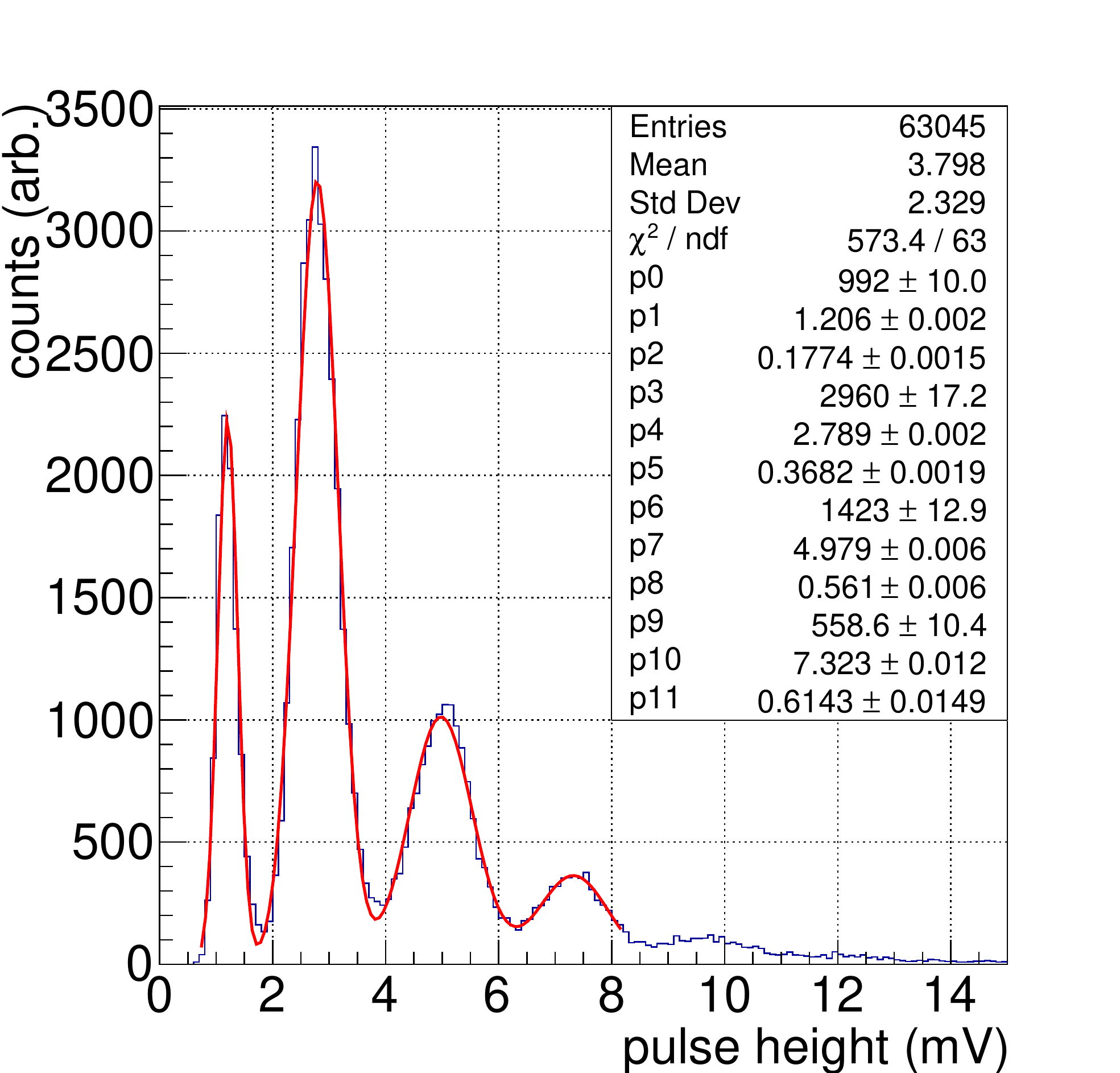}
		\caption{}
	\end{subfigure}
\caption{The response of the SiPMs to low photon counts for (a) channel 1 and (b) channel 2, with a four-Gaussian fit: the parameters are the amplitude, mean, and standard deviation for the four Gaussian distributions. The first peak is due to noise, and the subsequent three are the single, double and triple photon response. We calculate the gain as the average distance between the mean of the peaks: 0.226~mV for channel 1 and 0.223~mV for channel 2. Note that these data were acquired with a x10 amplifier, and the remaining bar characterizations were not. Data shown is for the SNL setup, though this process was repeated at UH with comparable results, as evident from Table~\ref{tab:spe}.}
\label{fig:fig3} 
\end{figure}

With both experimental setups, the single photon response of SiPMs used for each end of the 
bar was characterized so that an absolute comparison of the overall collection efficiency can be made 
between bars. In order to observe the single photon response, the FOUT was amplified 
with a Photek PA200-10 (Mini-Circuits ZFL-1000LN+) 20~dB amplifier at SNL (UH) and data was acquired with a random trigger and no light 
source in the enclosure. For two SiPM pixels at SNL, Figure~\ref{fig:fig3} shows the maximum pulse height for interactions in which a 
pulse was found over threshold. The first peak is due to noise pulses above the software threshold of 
the peak-finding algorithm, and the subsequent peaks are the single, double, and triple photon 
responses. The results of a four-gaussian fit are included. We calculate the single photon amplitude as the average distance 
between the mean of the real (not noise) peaks: 0.226~mV for channel 1 and 0.223~mV for channel 2 when 
corrected for the gain of the amplifiers. Table~\ref{tab:spe} lists the single photon pulse heights for both apparatuses.

\begin{table}
\centering
\begin{tabular}{|c|cc|}
\hline
Apparatus				& Channel 1 (mV)	& 	Channel 2 (mV)	\\
\hline\hline
UH					&0.2312			&0.2418\\
SNL					&0.2257			&0.2230\\
\hline
\end{tabular}
\caption{The single photon pulse height for the individual SiPM readout channels.}
\label{tab:spe}
\end{table}

\begin{figure}[!htbp]
		\centering
		\includegraphics[width=0.49\columnwidth]{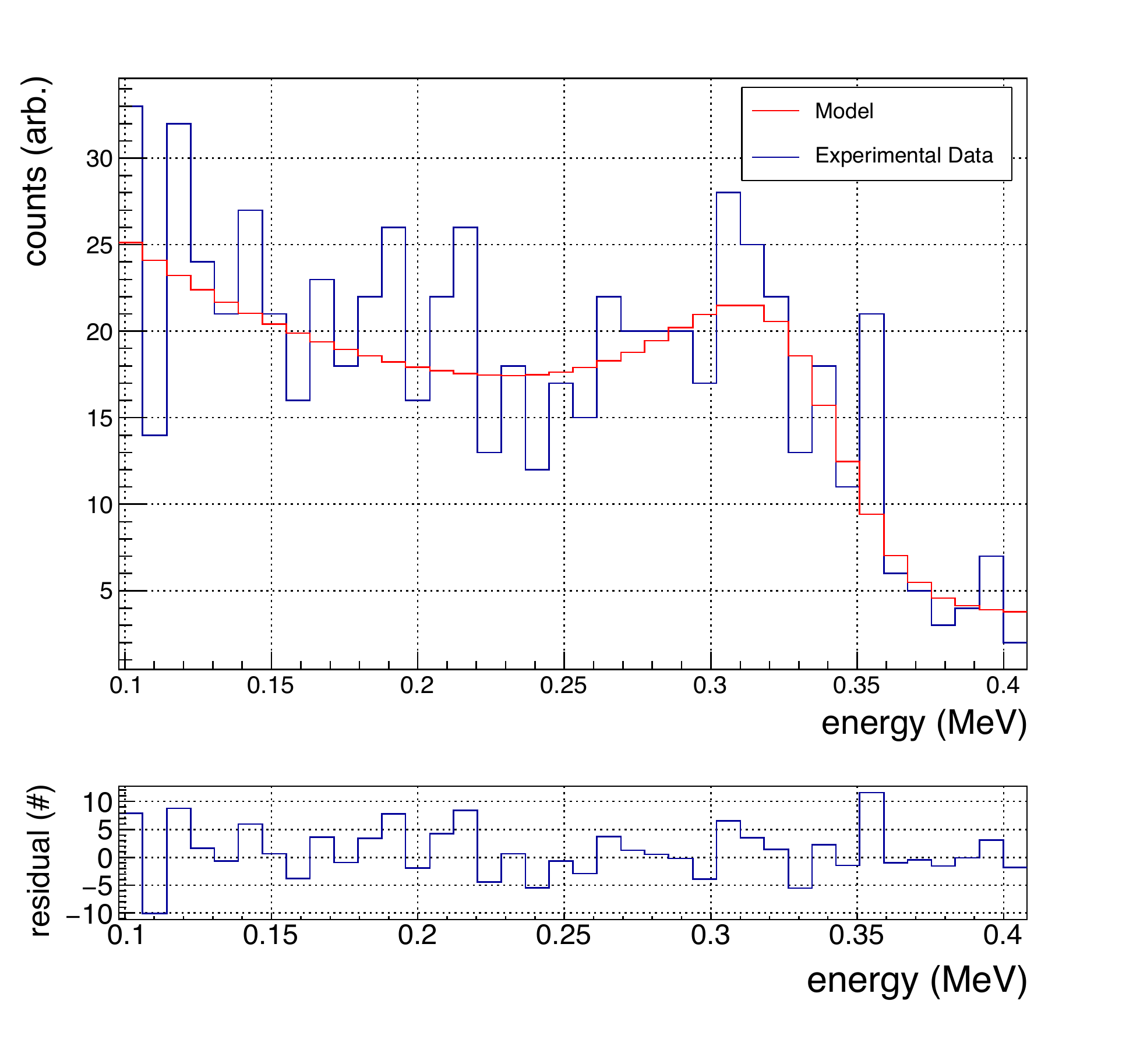}
	\caption{An example energy calibration using the coincident $^{22}$Na spectrum for the EJ-204 bar with no reflector. }
\label{fig:fig4} 
\end{figure}

For energy resolution measurements with $^{22}$Na tagged or $^{137}$Cs data, interactions
in the central 5~mm of the bar are selected by 
coincidence with the trigger scintillator ($^{22}$Na) or by using the timing difference calibration from the position scans ($^{137}$Cs). 
The resulting energy spectrum, calculated from the geometric mean of the calibrated energy measurements at each end of the bar, $E = \sqrt{E_1E_2}$, is used 
to compare to the expected spectrum without any resolution effects from the Klein-Nishina formula. In 
order to obtain the energy resolution, the experimental result is converted to keVee units with a linear 
transformation:
\begin{linenomath*}
\begin{equation}
E_{keVee} = q_0E_{mV}+q_1.
\end{equation}
\end{linenomath*}
Next, an arbitrary y-scale, $q_2$, is applied to the Klein-Nishina prediction for the gamma emissions resulting from the source superposed
with a power law contribution to account for down-scattered gammas, and a gaussian 
convolution is performed with a standard deviation of:
\begin{linenomath*}
\begin{equation}
\frac{\sigma}{E_{keVee}}= \sqrt{q_3^2 + \frac{q_4^2}{E_{keVee}} + \frac{q_5^2}{E_{keVee}^2} }.
\label{eq:res}
\end{equation}
\end{linenomath*}
The terms in Equation \ref{eq:res} represent geometrical effects of light 
transmission ($q_3$), the statistical variation of the production and 
multiplication of photo-electrons ($q_4$), and the electronic noise 
($q_5$)~\cite{klein}. The power law has an additional two parameters ($q_6$) and ($q_7$).
The minimization routine MINUIT~\cite{minuit1,minuit2} is used to find the 
parameters, $q_0$ through $q_7$, for which the $\chi^2$ of the convolved Klein-Nishina 
spectrum and measured energy spectrum is minimized.  This is a large 
parameter space, so many of the parameters are restricted in range, or in the 
example of the offset in the linear transformation, fixed to the known value.  The range over which the 
fit is performed is also restricted to the region of the Compton edge feature.
As an example, for a bare EJ-204 bar Figure~\ref{fig:fig4} 
shows the resulting convolved model prediction and calibrated experimental spectra after the minimization for the coincident $^{22}$Na spectrum.
Energy resolution results using both $^{22}$Na and $^{137}$Cs data for all
configurations are summarized in Table~\ref{tab:summary}.

\subsection{Position Response}
The position resolution is evaluated with both setups. It should be noted that, to increase performance by a factor of
ten compared to traditional two-plane scatter camera systems, we desire an overall position resolution of roughly 10~mm ($\sigma$)~\cite{svsc}. 
This translates to a z-position resolution of 9.8~mm, given that the $x$ and $y$ positions are $5/\sqrt{12}$~mm ($\sigma$).

Results are presented below for the position resolution evaluated with the difference in time of arrival at each end of the bar,
denoted $\sigma_z^t$, as well as the log of the charge amplitude ratio at the end of each bar, denoted $\sigma_z^A$. For each interaction, 
the difference in times and the log of the charge amplitude ratio are evaluated, and 
the distribution of each for a given scan position/interaction location is fit to a Gaussian function. Under the assumption of a single exponential attenuation model, the form of the log of the charge amplitude ratio with 
interaction location ($z$) is expected to be linear: 
\begin{linenomath*}
\begin{eqnarray}
\mathrm{ln}\frac{A_1}{A_2} &=& \mathrm{ln}\frac{e^{-z/\lambda}}{e^{-(L-z)/\lambda}} \nonumber \\
					&=& \frac{L}{\lambda} - \frac{2z}{\lambda},
\label{eq:7}					
\end{eqnarray}  
\end{linenomath*}
where $A_1/A_2$ are the pulse heights of channel 1 and 2, $z$ is the position along the bar, $L$ is the total length of the bar, and $\lambda$ is the effective attenuation length.
However, this assumes that the effective attenuation length (a function of reflection losses and the bulk attenuation) is the only source
of light loss and is constant throughout the scintillator bar. The difference in time is also expected to be linear if the velocity of light 
within the scintillator bar is constant throughout:
\begin{linenomath*}
\begin{eqnarray}
t_1-t_2 &=& \frac{z}{v} - \frac{L-z}{v} \nonumber \\
	&=& \frac{2z}{v} - \frac{L}{v}.
\label{eq:8}
\end{eqnarray}
\end{linenomath*}
The values are then combined to form the Best Linear Unbiased Estimator (BLUE)~\cite{blue}.

\subsubsection{Scanning Results with $^{90}$Sr and $^{22}$Na}
\begin{figure}[!htbp]
		\centering
		\begin{subfigure}[h]{0.49\columnwidth} 
		\includegraphics[width=\columnwidth]{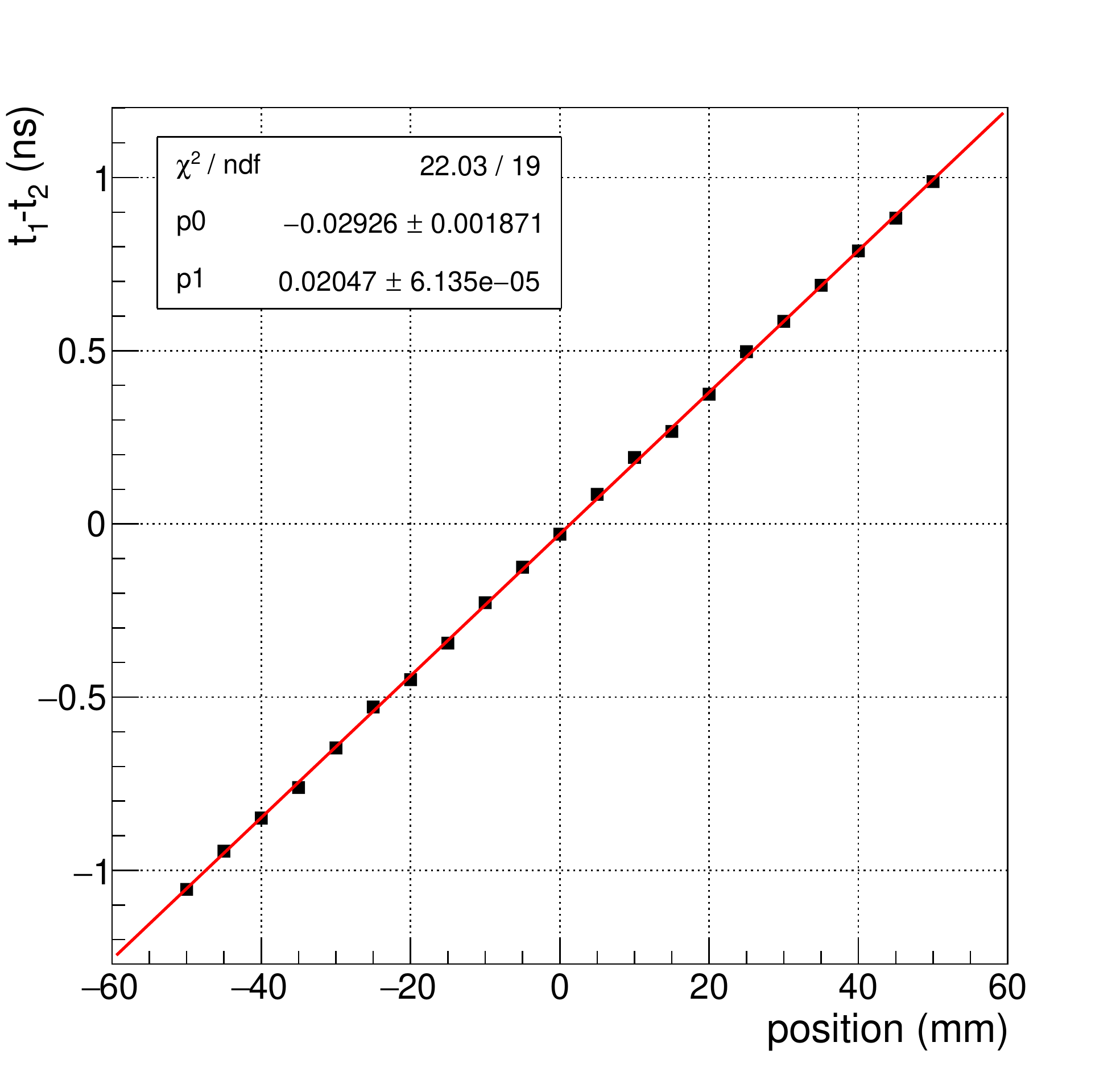}
		\caption{}
		\end{subfigure}
		\begin{subfigure}[h]{0.49\columnwidth} 
		\includegraphics[width=\columnwidth]{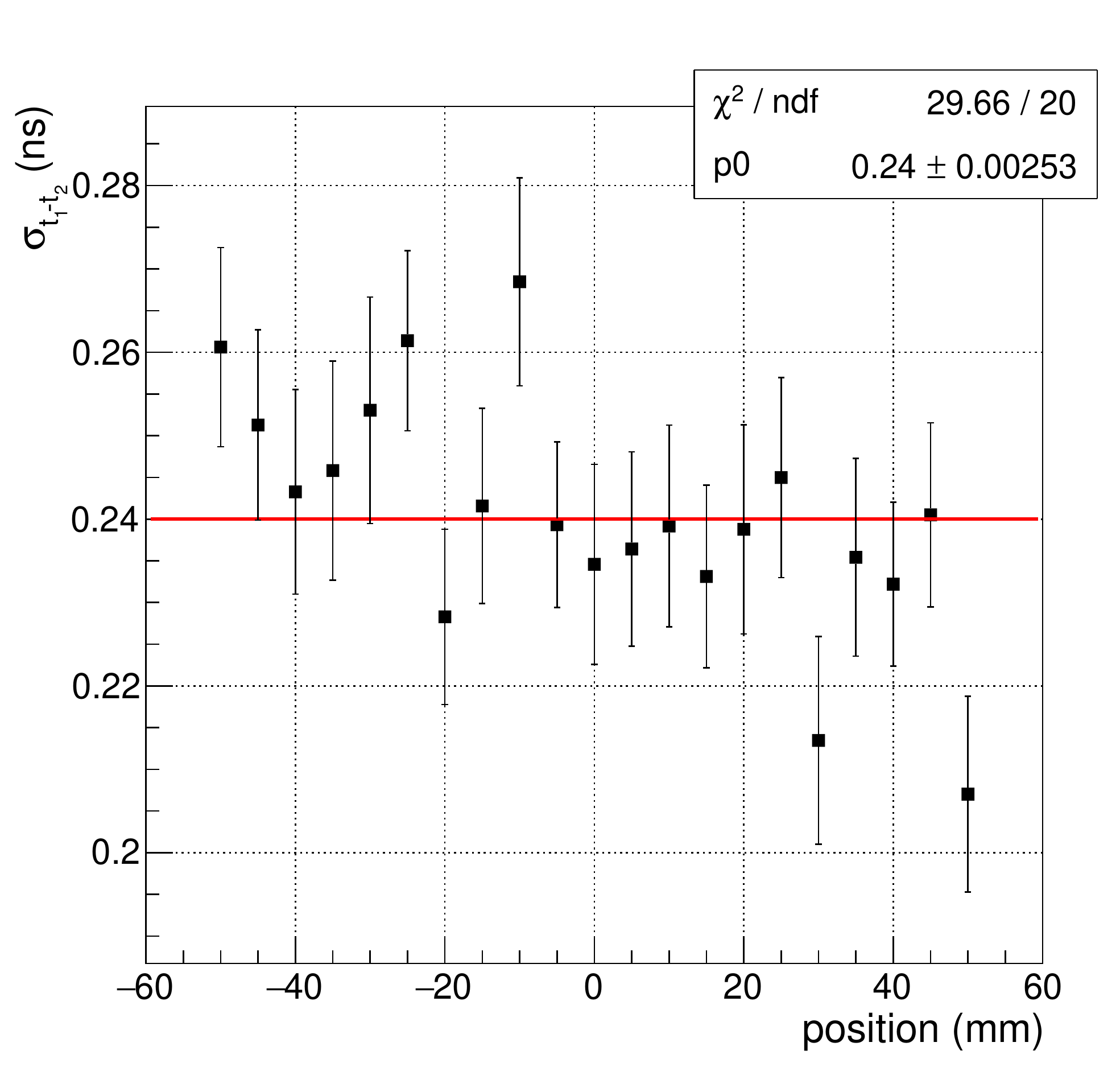}
		\caption{}
		\end{subfigure}
			\caption{(a) The difference in time of arrival between the two SiPMs as a function of $z$ and 
			(b) the standard deviation of the timing difference as a function of $z$. Both plots are for the EJ-204, 
			bare 1 run from the UH setup. Only energy depositions between 300-400 keVee are included.}
\label{fig:fig5} 
\end{figure}

\begin{figure}[!htbp]
		\centering
		\begin{subfigure}[h]{0.49\columnwidth} 
		\includegraphics[width=\columnwidth]{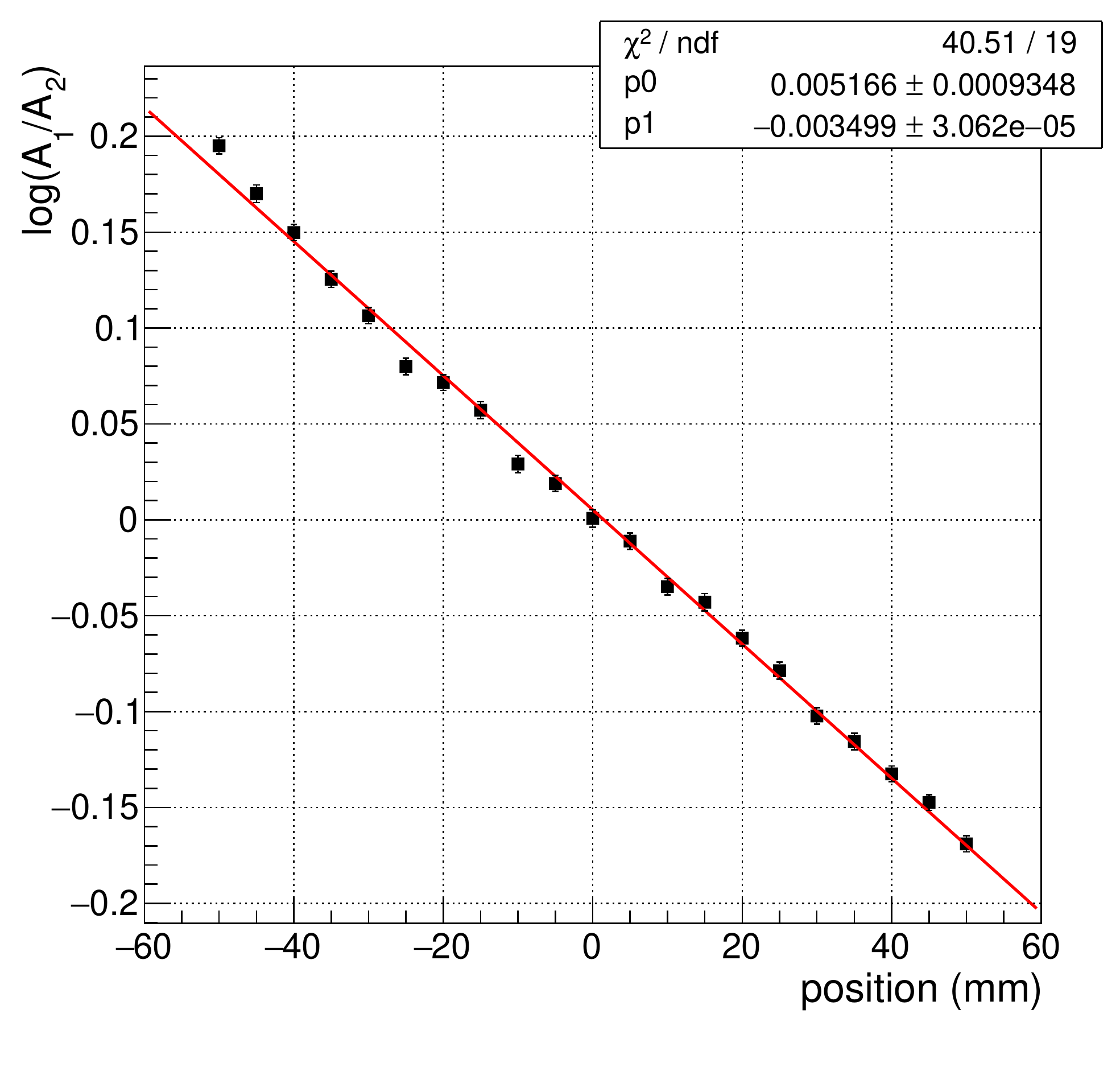}
		\caption{}
		\end{subfigure}
		\begin{subfigure}[h]{0.49\columnwidth} 
		\includegraphics[width=\columnwidth]{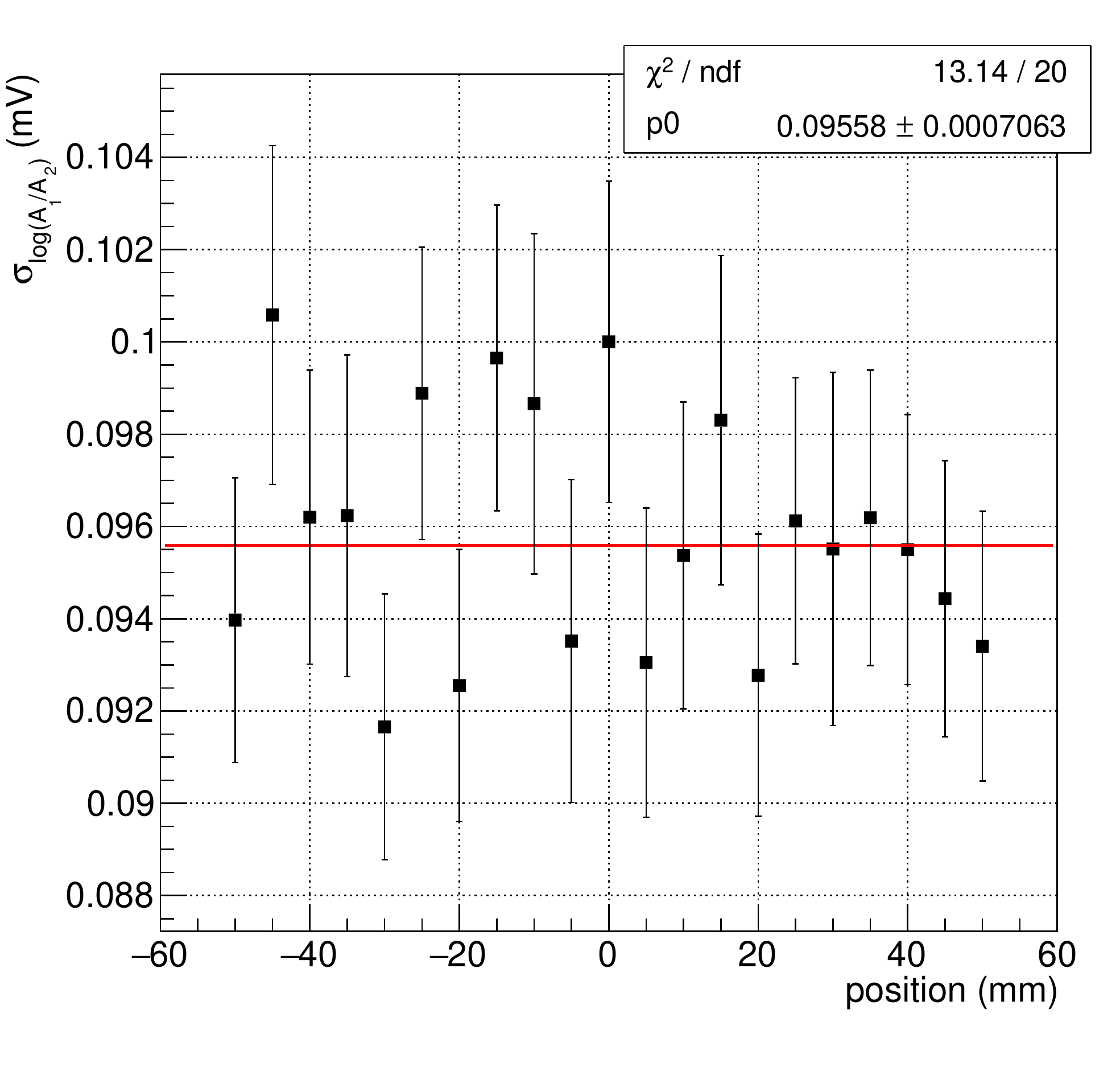}
		\caption{}
		\end{subfigure}
			\caption{(a) The log ratio of the charge amplitude as a function of interaction position and 
			(b) the standard deviation of the log ratio as a function of interaction position. Both plots are 
			for the EJ-204, bare 1 run from UH setup. Only energy depositions between 300-400 keVee are included.}
\label{fig:fig6} 
\end{figure}

\begin{figure}[!htbp]
		\centering
		\begin{subfigure}[h]{0.49\columnwidth} 
		\includegraphics[width=\columnwidth]{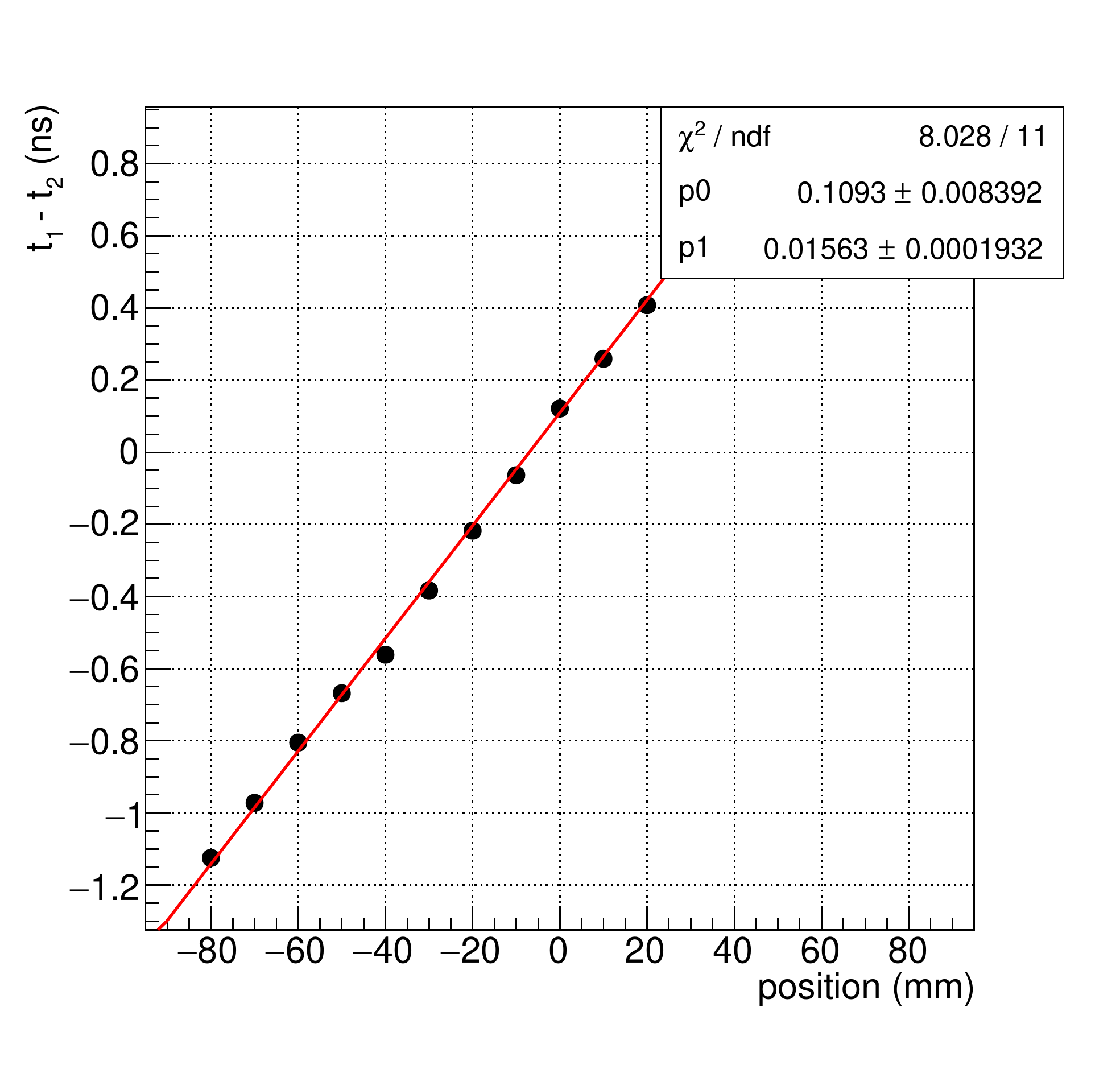}
		\caption{}
		\end{subfigure}
		\begin{subfigure}[h]{0.49\columnwidth} 
		\includegraphics[width=\columnwidth]{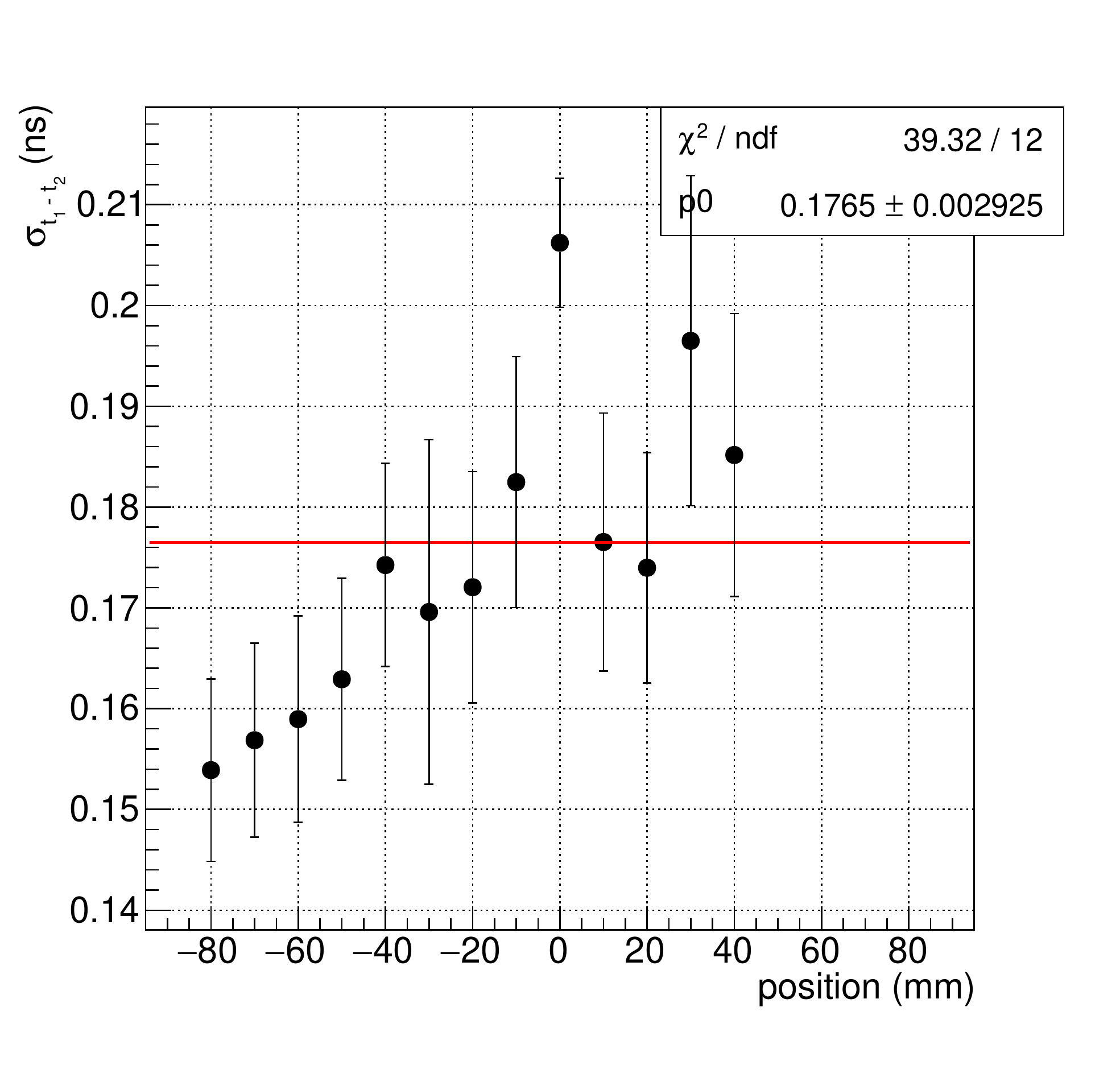}
		\caption{}
		\end{subfigure}
			\caption{(a) The difference in time of arrival between the two readout ends of the bar 
			as a function of $z$.  The fit parameters are for a first order polynomial, and the value on the plot is the position resolution. 
			(b) The standard deviation of the timing difference as a function of $z$. The fit parameters are for a zeroth order polynomial. 
			Both plots are for EJ-204 with no reflector. Only energy depositions between 300-400 keVee are included.}
\label{fig:fig7} 
\end{figure}

\begin{figure}[!htbp]
		\centering
		\begin{subfigure}[h]{0.49\columnwidth} 
		\includegraphics[width=\columnwidth]{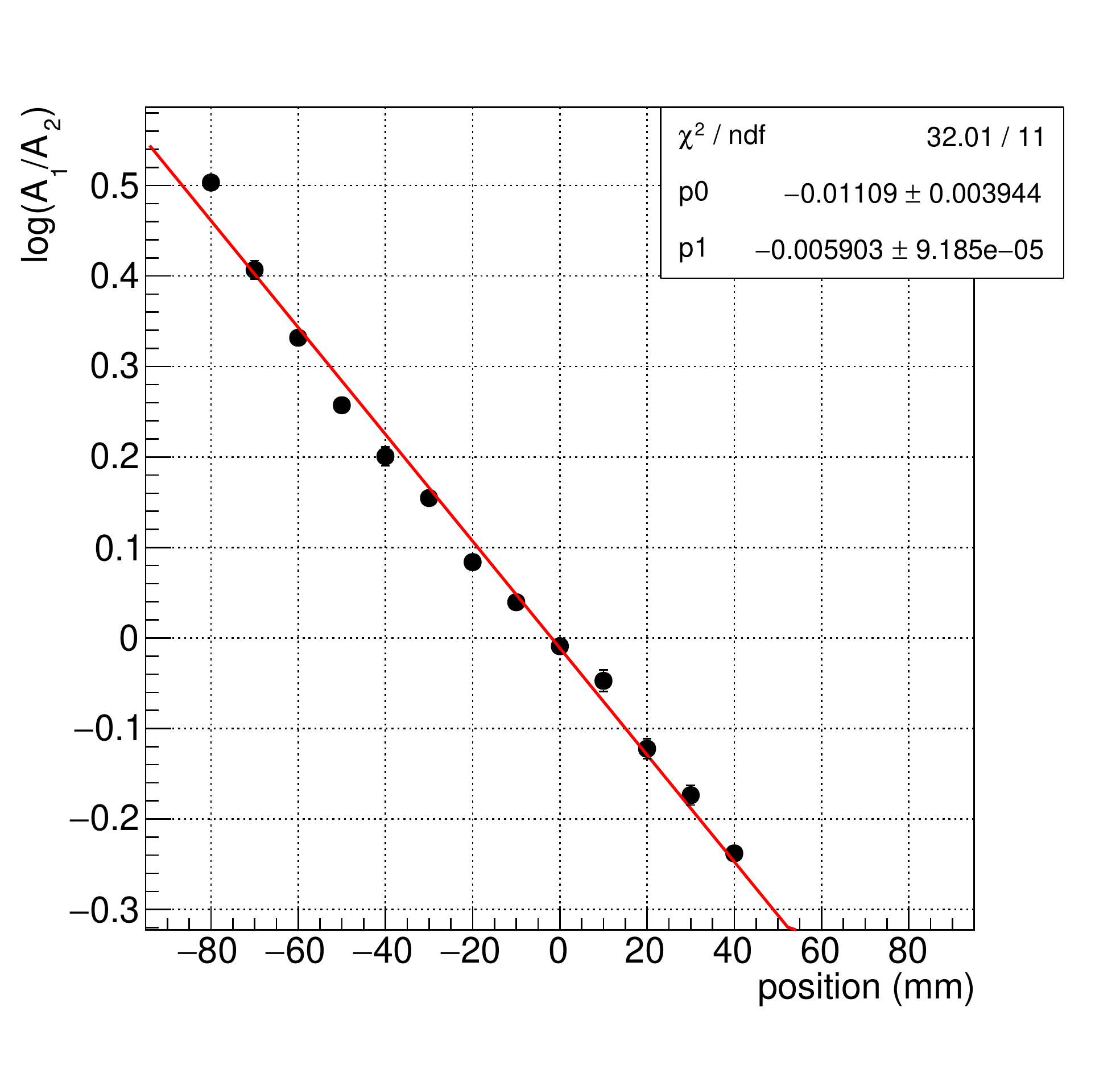}
		\caption{}
		\end{subfigure}
		\begin{subfigure}[h]{0.49\columnwidth} 
		\includegraphics[width=\columnwidth]{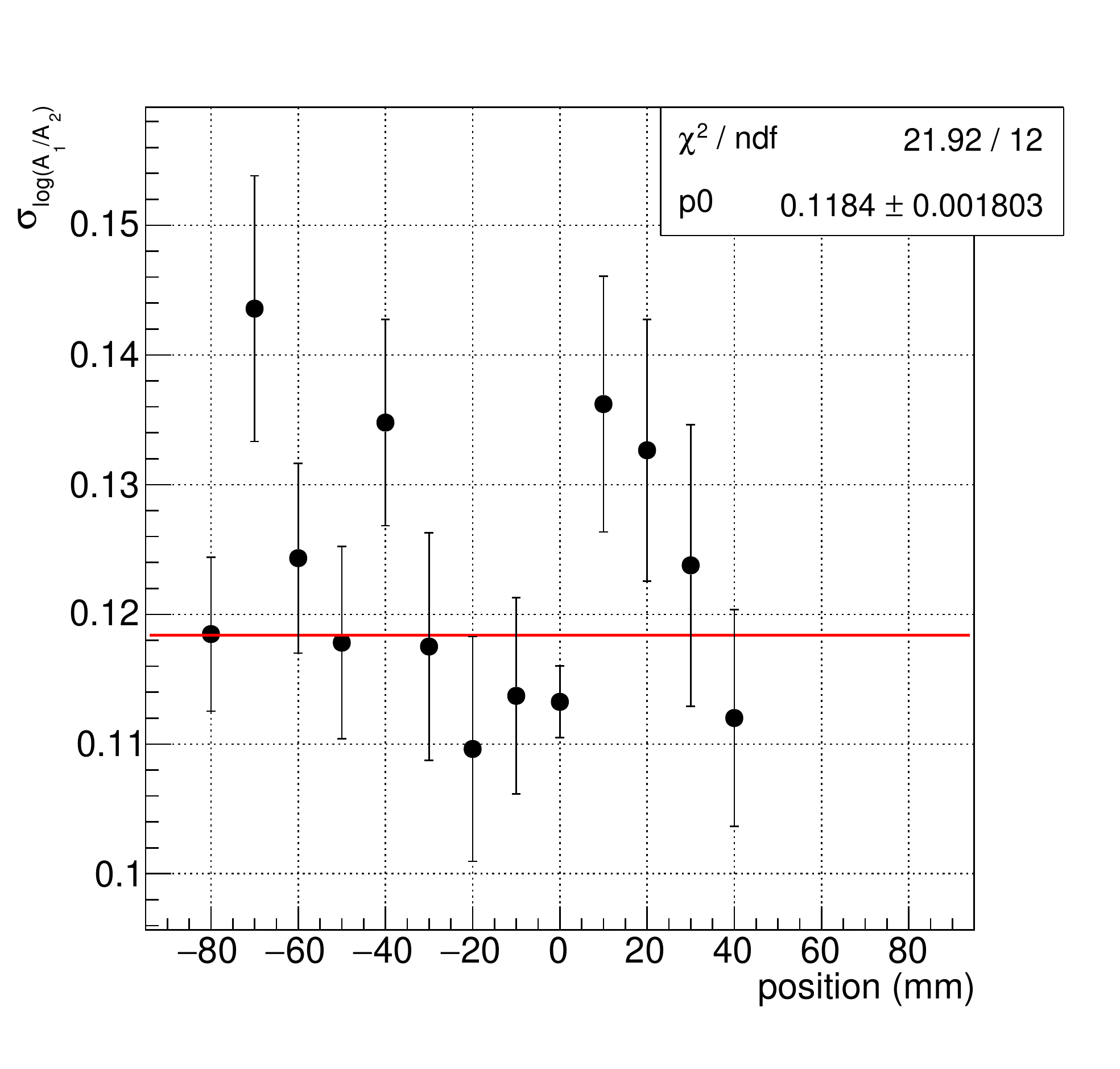}
		\caption{}
		\end{subfigure}
			\caption{(a) The log ratio of the charge amplitude as a function of $z$. 
			The fit parameters are for a first order polynomial, and the value on the plot is the position resolution.  
			(b) The standard deviation of the log ratio function of $z$. The fit parameters 
			are for a zeroth order polynomial. Both plots are for EJ-204 with no reflector. Only energy depositions between 300-400 keVee are included.}
\label{fig:fig8} 
\end{figure}

Example position characterizations with both the difference in timing and the log ratio of the charge amplitude
 is shown for $^{90}$Sr (Figures~\ref{fig:fig5} and \ref{fig:fig6}) and $^{22}$Na (Figures~\ref{fig:fig7} and \ref{fig:fig8}). 
Only interactions with an energy deposition between 300 and 400~keVee are used for this analysis. Both the mean of the 
 Gaussian fit and the standard deviation as a function of distance is shown, and the position resolution is calculated as $
 \frac{\sigma_{avg}}{m}$, in which $m$ is the slope of the first order polynomial fit to the mean 
 as a function of $z$, and $\sigma$ is the average of the standard deviations as a function of $z$ 
 determined with a zeroth order polynomial fit ($^{22}$Na) or as a simple average ($^{90}$Sr). 
Our data show only minor deviations from a linear fit in this regime.  For example, for the $^{22}$Na data, the maximum deviation from the linear fit on the order of 50 ps. 
 A summary of all the position resolution results are presented in Tables~\ref{tab:UHpos} and \ref{tab:SNLpos}. 
For the $^{22}$Na data, each measurement in Table~\ref{tab:SNLpos} corresponds to a specific and unique sample of scintillator, with the indicated wrapping.
In order to study systematic effects, repeated measurements were performed on some bars for the $^{90}$Sr scans, and bars used with wrappings were also tested bare.
 The errors quoted are statistical, propagated from the errors on the fit values. The BLUE combination of the 
first two columns is shown in the third.

\begin{table}\begin{center}
\begin{tabular}{|r||c|c||c||}
\hline
Scintillator		&$\sigma_z^{t}$ (mm)			&$\sigma_z^{A}$ (mm) 			& $\sigma_z$ (mm)		 \\	
\hline
EJ-200,  bar 1:	        & 16.27$\pm$0.18	 &	40.77$\pm$0.60	& 14.29	\\
bar 2:			& 16.14$\pm$0.19 	 &	39.51$\pm$0.58	& 13.99	\\
bar 3:			& 16.61$\pm$0.13 	 &	42.90$\pm$0.66	& 14.53	\\ 
Teflon bar 3:			& 13.31$\pm$0.18	 &	10.37$\pm$0.10&  7.65	\\ 
ESR, bar 1:		& 14.07$\pm$0.17	 & 	35.58$\pm$0.51  & 12.31	\\ 
bar 2:			& 13.94$\pm$0.17	 & 	29.29$\pm$0.37  & 11.87 	\\
\hline
EJ-204, bar 1:	        & 12.37$\pm$0.14	 &	33.46$\pm$0.44	& 10.77	\\
bar 2:			& 12.44$\pm$0.15	 &	36.32$\pm$0.53	& 10.86	\\
bar 3:			& 11.73$\pm$0.19 	 &	27.32$\pm$0.31	& 10.03    \\ 
bar 4:			& 12.28$\pm$0.14 	 &	40.57$\pm$0.63	& 11.03    \\ 
Teflon, bar 4:		 & 10.89$\pm$0.14	 &	 9.58$\pm$0.08	&  6.54	\\
ESR, bar 1:	        & 10.43$\pm$0.12	 &	22.08$\pm$0.24	&  8.74	\\ 
bar 1:			& 11.16$\pm$0.15	 &	36.68$\pm$0.57  & 10.13	\\
bar 1:			& 11.31$\pm$0.14	 &	34.85$\pm$0.52	& 10.15	\\
bar 2:			& 11.05$\pm$0.14	 &	28.25$\pm$0.35  &  9.53	\\ 
\hline
EJ-230, bar 1:	        & 10.80$\pm$0.14	 &	22.79$\pm$0.27	&  8.85	\\ 
bar 2:			& 10.93$\pm$0.14	 &	22.44$\pm$0.26	&  8.74	\\ 
bar 3:			& 10.51$\pm$0.13 	 &	25.52$\pm$0.32	&  8.91	\\ 
bar 4:			& 10.69$\pm$0.13	 &	24.40$\pm$0.30	&  8.94	\\ 
Teflon, bar 1:		& 10.88$\pm$0.32	 &	 9.30$\pm$0.23	&  6.32	\\ 
ESR, bar 1:		& 10.04$\pm$0.11	 &	23.57$\pm$0.25	&  8.43	\\
bar 1:			& 10.35$\pm$0.11	 &	24.76$\pm$0.26	&  8.59	\\
bar 2:			& 10.35$\pm$0.12	 & 	23.32$\pm$0.24	&  8.53	\\ 
\hline		
EJ-276, bar 1:          & 18.38$\pm$0.25         &  17.49$\pm$0.20  &  13.17 \\ 
bar 2:              		   & 19.07$\pm$0.27         &  18.56$\pm$0.22  &  13.85 \\ 
Teflon, bar 1:                 & 16.66$\pm$0.26         &  10.83$\pm$0.14  &   9.54 \\ 
ESR, bar 1:                    & 13.76$\pm$0.19         &  17.18$\pm$0.20  &  10.45 \\
\hline  
\end{tabular}
\caption{Summary of the position resolution results of UH. The errors are statistical only, on the fit of the particular distribution; see Table \ref{tab:UH-SNL-compare} for an estimate of the systematic errors. The third column is the best linear unbiased estimator of the first two columns.}
\label{tab:UHpos}
\end{center}
\end{table}

\begin{table}\begin{center}
\begin{tabular}{|rr||c|c||c||}
\hline
\multicolumn{2}{|c||}{Scintillator}			&$\sigma_z^{t}$ (mm)	&$\sigma_z^{A}$ (mm) 	& $\sigma_z$ (mm)		 \\	
\hline
\multicolumn{2}{|l||}{EJ-200, bare:}			&14.58$\pm$0.26        		&33.10$\pm$0.64     		&13.34	\\
				& bare:				&14.74$\pm$0.36        		&31.55$\pm$0.87      	&13.35\\
				& Teflon:				&14.17$\pm$0.36       		&14.96$\pm$0.39      	&10.29\\
				& ESR:				&12.82$\pm$0.28        		&22.52$\pm$0.54      	&11.14	\\
				\hline
\multicolumn{2}{|l||}{EJ-204, bare:}			&11.93$\pm$0.28       		&20.59$\pm$0.53    		&10.32\\
				& bare:				&11.29$\pm$0.23        		&20.06$\pm$0.44      	& 9.84\\
				& Teflon:				&12.33$\pm$0.26       		&10.66$\pm$0.22      	& 8.06\\
				& ESR:				& 9.83$\pm$0.26    	   		&17.69$\pm$0.50		& 8.59\\
				\hline
\multicolumn{2}{|l||}{EJ-230, bare:}			&11.31$\pm$0.22      		&18.24$\pm$0.38      	& 9.61\\
				& bare:				&11.46$\pm$0.26       		&19.02$\pm$0.47      	& 9.82\\
				& Teflon:				&11.23$\pm$0.24       		&12.62$\pm$0.28      	& 8.39\\
				& ESR:				&12.30$\pm$0.30        		&18.10$\pm$0.46      	&10.17\\
\hline
\multicolumn{2}{|l||}{EJ-276, bare:}			&17.83$\pm$0.59        		&18.95$\pm$0.72      	&12.98\\
				& bare:				&16.56$\pm$0.46       		&15.37$\pm$0.40      	&11.27\\
				& Teflon:				&15.41$\pm$0.32        		&11.65$\pm$0.24      	& 9.29\\
				& ESR:				&15.45$\pm$0.46       		&17.74$\pm$0.59     		&11.65\\
\hline
\end{tabular}
\caption{Summary of position resolution results from the SNL setup. The errors are statistical only on the fit of the particular distribution; see Table \ref{tab:UH-SNL-compare} for an estimate of the systematic errors. The third column is the best linear unbiased estimator of the first two columns. }
\label{tab:SNLpos}
\end{center}
\end{table}

\subsubsection{Comparison}

The two independent measurement setups and sources provide a valuable tool to compare results and identify significant systematic effects that can cause differences in effective resolutions. 
An understanding of such systematic effects can lead to component selection and design decisions that improve the robustness of the resolution results.
In order to make a direct comparison for each scintillator and wrapping at each test site, we have collected multiple repeated measurements into a combined measurement for each setup.
This comparison is shown in Table~\ref{tab:UH-SNL-compare}, which accumulates results based off of entries from Tables~\ref{tab:UHpos} and \ref{tab:SNLpos}. 

\begin{figure}[!htbp]
	\centering
	\includegraphics[width=\columnwidth]{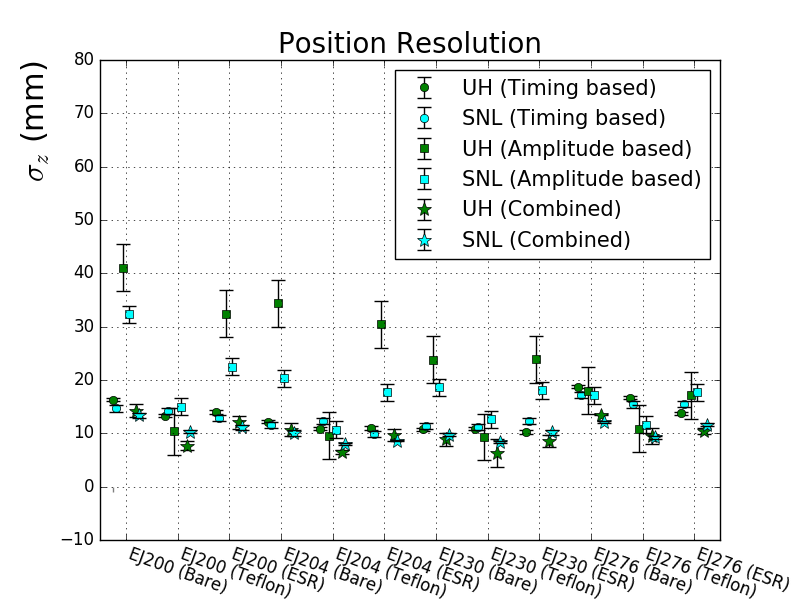}
	\caption{Position resolution results summarized for the two experimental
setups, using the timing information, amplitude information, and the BLUE
combination. Error bars reflect systematic uncertainties only, as reported in
Table~\ref{tab:UH-SNL-compare} and described in detail in the text, since these
dominate the uncertainties.}
\label{fig:fig9} 
\end{figure}

When multiple measurements are made for different samples of the same scintillator and surface treatment, these measurements are averaged into a single entry.
Statistical uncertainties in Table~\ref{tab:UH-SNL-compare} are based on the average of statistical uncertainties for multiple entries, if available.
Systematic uncertainties dominate the measurements, as can be seen from the significant variations in timing-based and amplitude-based resolution both between sites as well as within measurements taken at the same site using the same configuration. Our systematic uncertainty estimation assumes that systematic uncertainties at each site are unique, but that there is a single systematic uncertainty at each site that dominates uncertainties for all measurements taken there.  We thus assume that our measurements at each site are normally distributed, with a standard deviation corresponding to the site's systematic uncertainty. In this case, repeated measurements of resolution for the same material with the same surface treatment can be considered as drawing two samples from a single normal distribution of resolutions. We can estimate a standard deviation of all such differences, and then infer a site's systematic uncertainty, $s$, using the following formula:
\begin{linenomath*}
\[s = \sqrt{\frac{1}{2}\frac{1}{N-1}\sum_{\mathrm{all\,pairs}} \left(\sigma_\mathrm{A} - \sigma_\mathrm{B}\right)^2},\]
\end{linenomath*}
where A and B correspond to a pair of repeated measurements for the same configuration, and $N$ is the total number of measurement pairs. A graphical depiction of the results, including these systematic uncertainties, is shown in Figure~\ref{fig:fig9}.

\begin{table*}
\begin{center}
\begin{tabular}{|r||c|c||c|c|}
\hline
	\multirow{2}{*}{Scintillator}	&\multicolumn{2}{c||}{$\sigma_z^t$ (mm)}&\multicolumn{2}{c|}{$\sigma_z^{A}$  (mm)} \\

\cline{2-5}
		        & [UH] 	 & [SNL]    & [UH]      & [SNL] \\
\hline
EJ-200, bare& 16.34$\pm$0.19   & 14.66$\pm$0.31    & 41.06$\pm$0.61   & 32.33$\pm$0.76\\
 	Teflon& 13.31$\pm$0.18 & 14.17$\pm$0.36     	& 10.37$\pm$0.09   & 14.96$\pm$0.39\\
	 ESR& 14.01$\pm$0.17  & 12.82$\pm$0.28	& 32.44$\pm$0.44   & 22.52$\pm$0.54\\
\hline
EJ-204, bare& 12.21$\pm$0.15 & 11.61$\pm$0.26     & 34.42$\pm$0.48   & 20.33$\pm$0.49\\
	 Teflon& 10.89$\pm$0.15 & 12.33$\pm$0.26      &  9.58$\pm$0.78   & 10.66$\pm$0.22\\
	 ESR& 10.99$\pm$0.14  & 9.83$\pm$0.26	& 30.47$\pm$0.42   & 17.69$\pm$0.50\\
\hline
EJ-230, bare& 10.73$\pm$0.18 & 11.39$\pm$0.24     & 23.79$\pm$0.29   & 18.63$\pm$0.43\\ 
	 Teflon& 10.88$\pm$0.82 & 11.23$\pm$0.24	  &  9.30$\pm$0.23   & 12.63$\pm$0.28\\	
	 ESR& 10.25$\pm$0.11 & 12.30$\pm$0.30    		& 23.88$\pm$0.25   & 18.01$\pm$0.26\\
\hline		
EJ-276, bare& 18.73$\pm$0.26 & 17.20$\pm$0.53  & 18.03$\pm$0.21   & 17.16$\pm$0.56\\ 
	Teflon& 16.66$\pm$0.26 & 15.41$\pm$0.32	  & 10.83$\pm$0.14   & 11.65$\pm$0.32\\
	 ESR& 13.76$\pm$0.19 & 15.45$\pm$0.46	   & 17.18$\pm$0.20   & 17.74$\pm$0.46\\
\hline  
Systematic error&$\pm$0.30	&$\pm$0.59	&$\pm$4.45	 &$\pm$1.64\\
\hline
\end{tabular}
\caption{Summary comparison of data from the $^{90}$Sr data taken at UH and the $^{22}$Na data taken at SNL. The errors for each entry are statistical, and the systematic errors are listed in the last row. Details on the calculation of these values and errors are given in the text.}
\label{tab:UH-SNL-compare}
\end{center}
\end{table*}

The summary comparisons presented in Table \ref{tab:UH-SNL-compare} and shown graphically in Figure~\ref{fig:fig9} show a few notable effects.
In general, using timing differences seems to be a more robust approach to calculating $z$: the agreement between the $^{90}$Sr and $^{22}$Na results is much closer for most measurements, despite significant differences in the experimental setups (to be described in Section~\ref{sec:systematics}).
Furthermore, the systematic uncertainties associated with the timing-based measurements are generally much smaller, indicating better repeatability between identical setups.

We note that there are multiple methods that could be used to estimate systematic uncertainties, in particular with the $^{90}$Sr data taken at UH, since there are many repeated measurements over identical configurations.  
It is clear from the $^{90}$Sr data in Table \ref{tab:UHpos} that the large systematic uncertainties are driven primarily by the EJ-204 measurements.  
If these are excluded from the above analysis, the amplitude-based systematic uncertainty drops from 4.5~mm to 1.8~mm, giving a result quite consistent between sites.  
Since the root cause of this larger variation for the EJ-204 samples is unclear, we include the EJ-204 data in the composite estimate to be conservative.  
We also note that the ESR data hints at larger systematics using ESR than that seen with bare bars.  
Again, it is not clear from the limited sample size whether this is a feature of the ESR wrapping or was specific to these samples or experimental setups, so we conservatively include these samples when calculating the overall site uncertainty.

\subsubsection{Studies on Systematic Variations between Sites}
\label{sec:systematics}
While the relative trends in $\sigma_{z}^{A}$ are the same in the two setups, the absolute
agreement for $\sigma_{z}^{A}$ is notably worse than for $\sigma_{z}^{t}$.
In addition, the systematic uncertainties on $\sigma_{z}^{A}$ are large, indicating substantial variations observed within data on the same setups.
A number of studies were conducted in order to better understand these variations in $\sigma_{z}^{A}$, all conducted using the $^{90}$Sr setup at UH, and using bare EJ-204 bars.

\begin{figure}[!htbp]
		\centering
		\includegraphics[width=0.5\columnwidth]{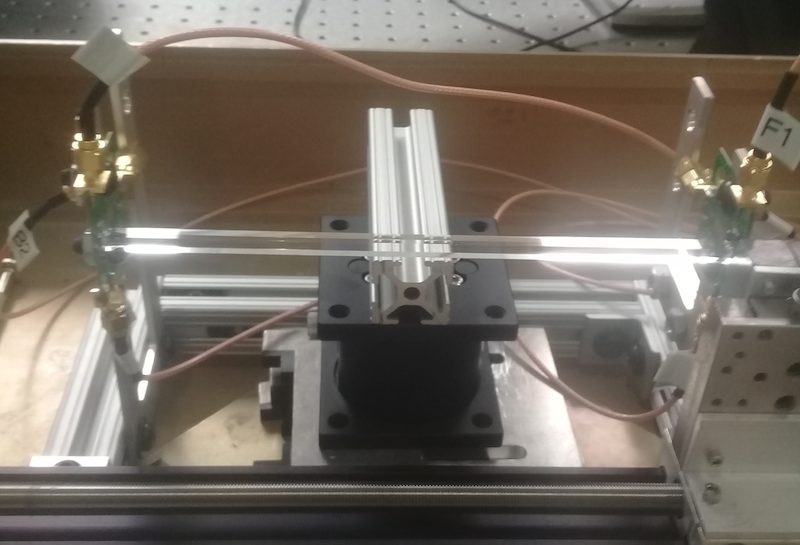}
	\caption{A picture of the test with the scintillator bar sustained by a support}
\label{fig:fig10} 
\end{figure}

The experimental setup for EJ-204 with no reflector material, which shows the most significant difference between the two measurement sites, was re-tested.
Three different scenarios based on possible systematic differences are investigated and summarized in Table \ref{tab:6}:
\begin{enumerate}
  \item Modified optical coupling, using approximately three times more than the typical amount of EJ-550 silicone grease.
  \item Modified triggering conditions, using a logical OR of trigger signals from either side of the bar, compared to the baseline AND trigger.
  \item Added a metal support underneath the EJ-204 bar near the midpoint between the two ends, as shown in Figure~\ref{fig:fig10}.
\end{enumerate}

\begin{table}
\centering
\begin{tabular}{|c||c|c|c||c|c|c|}
\hline
\multirow{2}{*}{Config.} &\multicolumn{3}{c||}{$t_1-t_2$}	&\multicolumn{3}{|c|}{$\mathrm{log}\frac{A_1}{A_2}$}\\
\cline{2-7}
				&$\sigma_z^t $(mm)   &$p_1$ (ns/mm)	&$p_0 $(ns)	&$\sigma_z^A$ (mm)&$p_1$(m$^{-1}$) &$p_0$\\
				\hline
		1               & 12.88$\pm$0.15 & 0.020 & 0.253 & 31.87$\pm$0.43 & -3.20 & 0.102 \\
		        2               & 13.17$\pm$0.11 & 0.020 & 0.267 & 29.74$\pm$0.26 & -3.47 & 0.103 \\
                        3               & 11.13$\pm$0.07 & 0.018 & 0.207 & 20.04$\pm$0.11 & -4.55 & 0.091 \\

\hline
\end{tabular}
\caption{The results of systematic studies in $\sigma_z^t$ and $\sigma_z^A$, including fit results used to calculate the resolutions in each configuration. See text for a description of the individual test configurations.}
\label{tab:6}
\end{table}

The position resolutions using the difference in time remained within the systematic errors, with the exception of the addition of the metal support (configuration 3). 
For the log ratio of the charge amplitude, an improvement of less than 5\% was obtained in $\sigma_{z}^{A}$ by adding more silicon grease. It rose up to 11.1\% when the condition of the trigger was changed, and it reached 40.1\% up when a support for the bar was placed.

The relative invariance of the timing-based results is generally as-expected, since the rising edge of the pules is less susceptible to photon variations compared amplitude measurements, which should vary proportionally to the total number of detected photons. The number of detected photons can vary with optical coupling, triggering, or interaction with support structures.
Optical coupling has a clear impact on observed amplitudes, so better coupling should generally improve performance due to a higher signal-to-noise ratio.
Triggering conditions can cause an asymmetry in the types of interactions observed: the standard AND based trigger can result in a bias toward generally larger depositions, whereas depositions resulting in a single large pulse on one end and a small pulse on the other end might be missed entirely.
These types of interactions are precisely those that would show good amplitude-based resolutions due to the large asymmetry between near- and far-end amplitudes, so this improvement is not unexpected.

On the other hand, the addition of the support structure had a much larger impact than originally anticipated.
The significant improvement in $\sigma_{z}^{A}$ is coupled with a substantially lower light yield, as observed in $^{137}$Cs calibration data.
This implies that the support itself is causing an overall decrease in light collection, of order $\sim$20\% for bare bars, possibly due to interfering with TIR at the contact positions. 
Since the baseline $^{90}$Sr data was taken with no mid-bar support and the $^{22}$Na data was taken with supports, this could explain some of the systematic differences seen there. 
It should be noted, however, that the support types likely differ in their effect on optical transport because of their shapes (rectangular vs. cylindrical) and reflectivity (bare aluminum vs. 
black felt wrapped). The effect of a support is less clear in the cases of ESR or Teflon, though we speculate that it should impact Teflon the least, since it is already in good contact 
with the bar, compared to ESR, which is loosely fitted around the bar.

\subsubsection{Performance as a Function of Deposited Energy}
\begin{figure}[!htbp]
		\centering
		\begin{subfigure}[h]{0.45\columnwidth} 
		\includegraphics[width=\columnwidth]{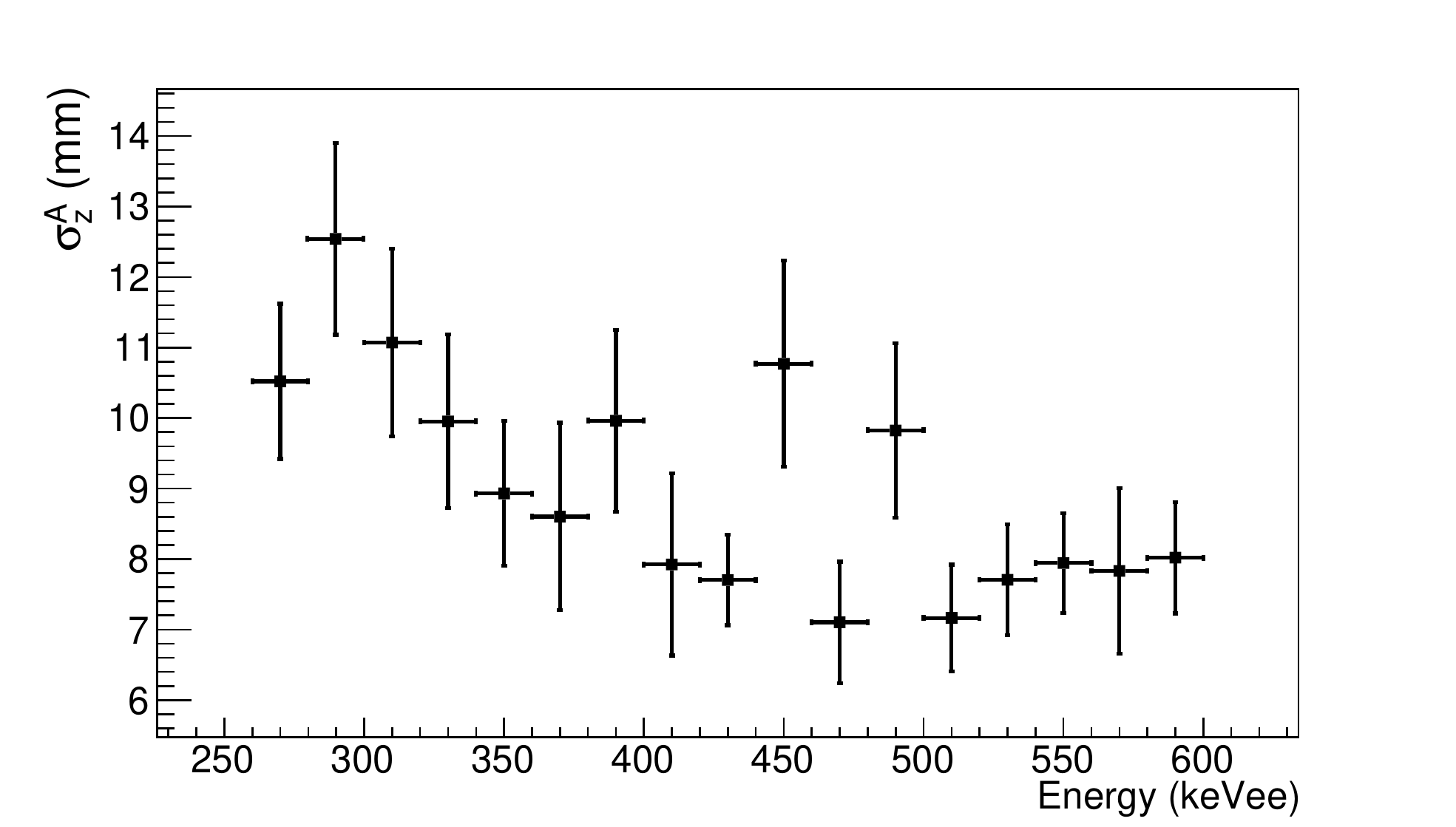}
		\caption{}
		\end{subfigure}
		\begin{subfigure}[h]{0.45\columnwidth} 
		\includegraphics[width=\columnwidth]{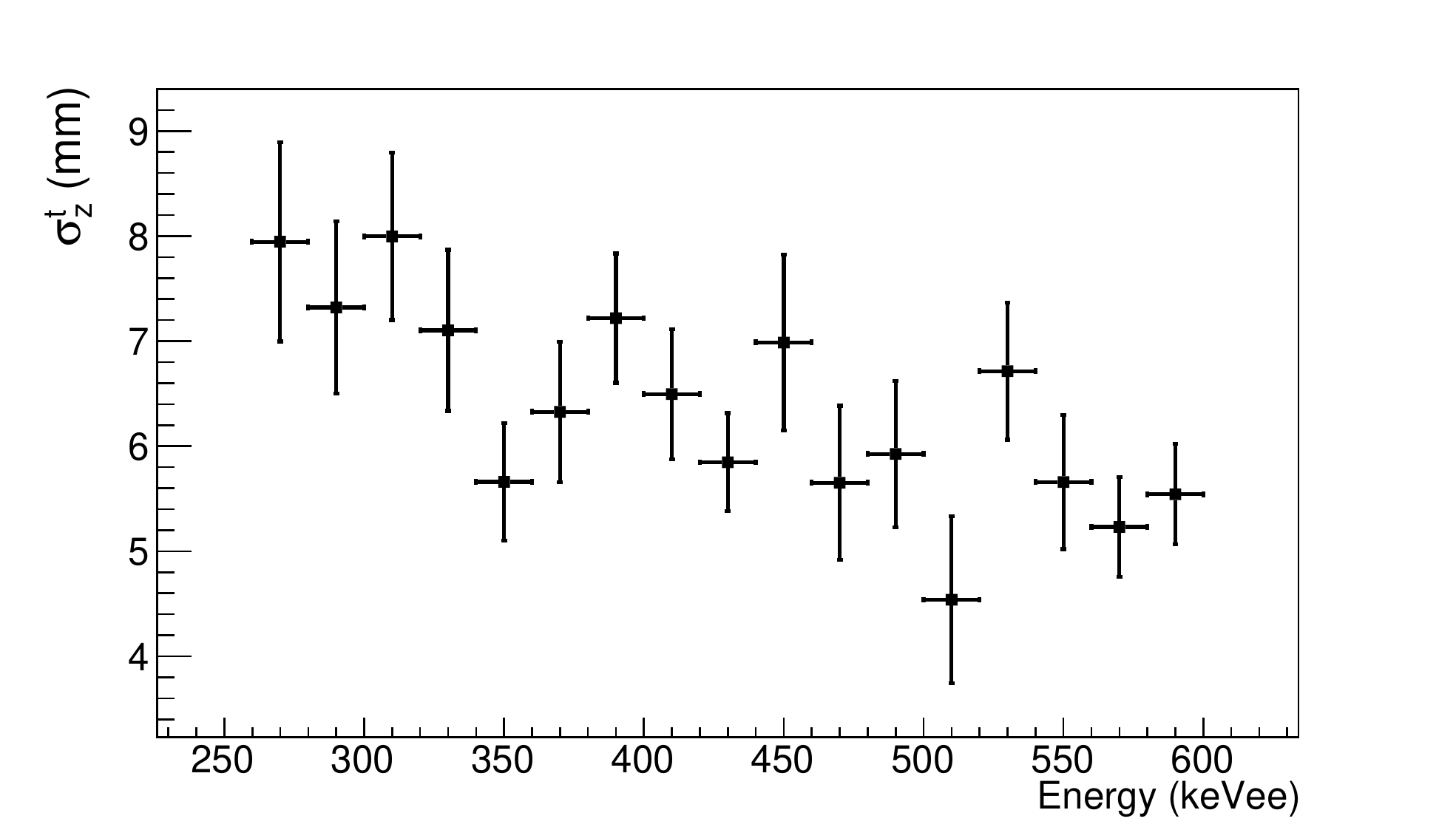}
		\caption{}
		\end{subfigure}
	\caption{Position resolution results using time differences (a) and log amplitude ratios (b) as a function of energy for Teflon wrapped EJ-204 using $^{90}$Sr data.}
\label{fig:fig11} 
\end{figure}

It is expected that there will be variation in position resolution as a function of deposited energy due to improved photostatistics, which are not readily apparent in the limited 300--400~keVee measurements reported to this point.  Figure~\ref{fig:fig11} shows $\sigma_z^t$ and $\sigma_z^A$ for Teflon-wrapped EJ-204 as a function of energy from the $^{90}$Sr data. 
Using the same position calibration as that obtained for the nominal 300--400~keVee bin, resolutions were studied in the range of 250--600~keVee.
As expected, both resolutions improve with energy, indicating that system performance for neutron imaging is expected to be better than our reported summary resolutions for higher energy depositions.

\subsection{Timing Response}
\begin{figure}[!htbp]
		\centering
		\begin{subfigure}[h]{0.29\columnwidth} 
		\includegraphics[width=\columnwidth]{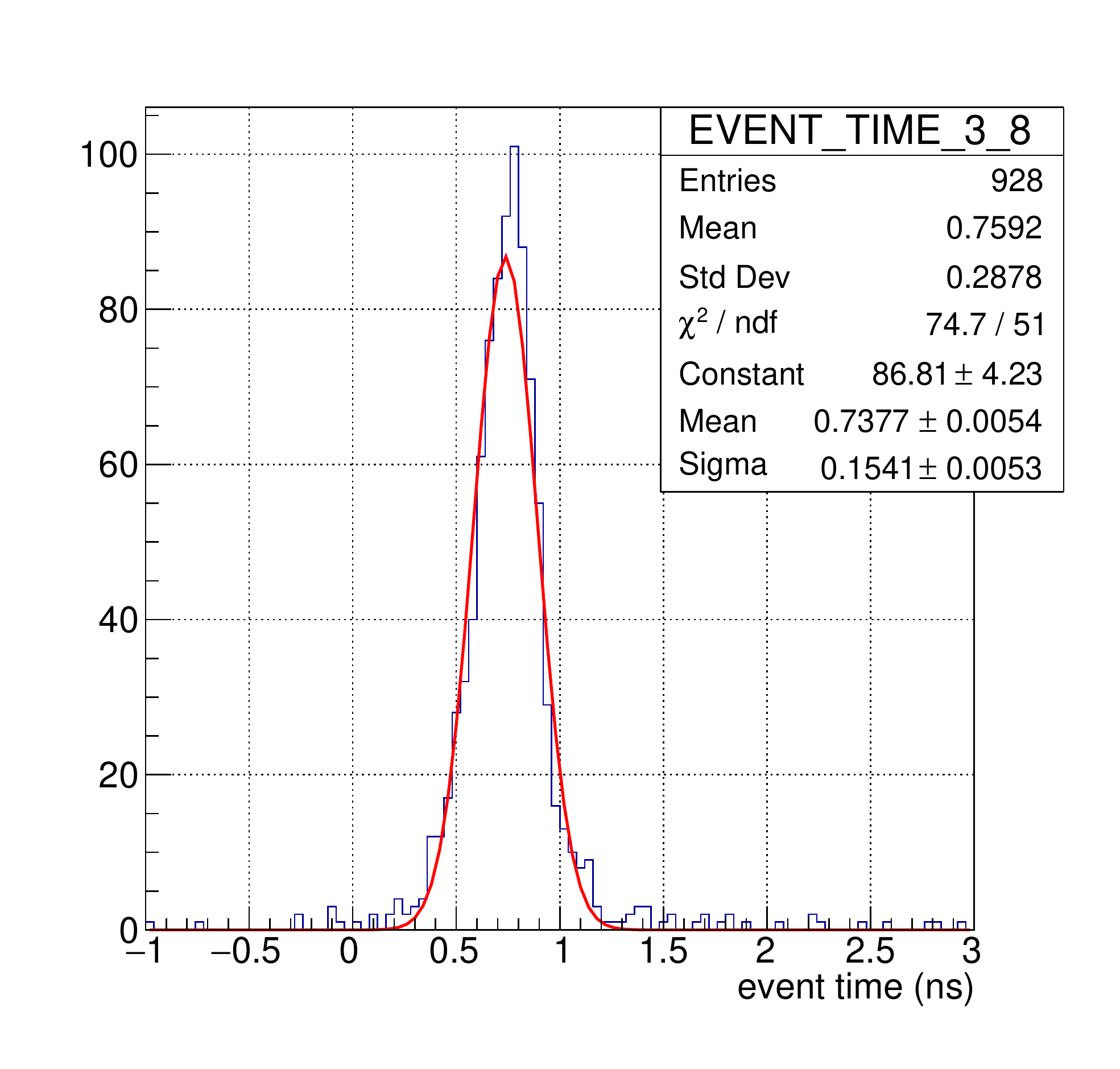}
		\caption{}
		\end{subfigure}
		\begin{subfigure}[h]{0.29\columnwidth} 
		\includegraphics[width=\columnwidth]{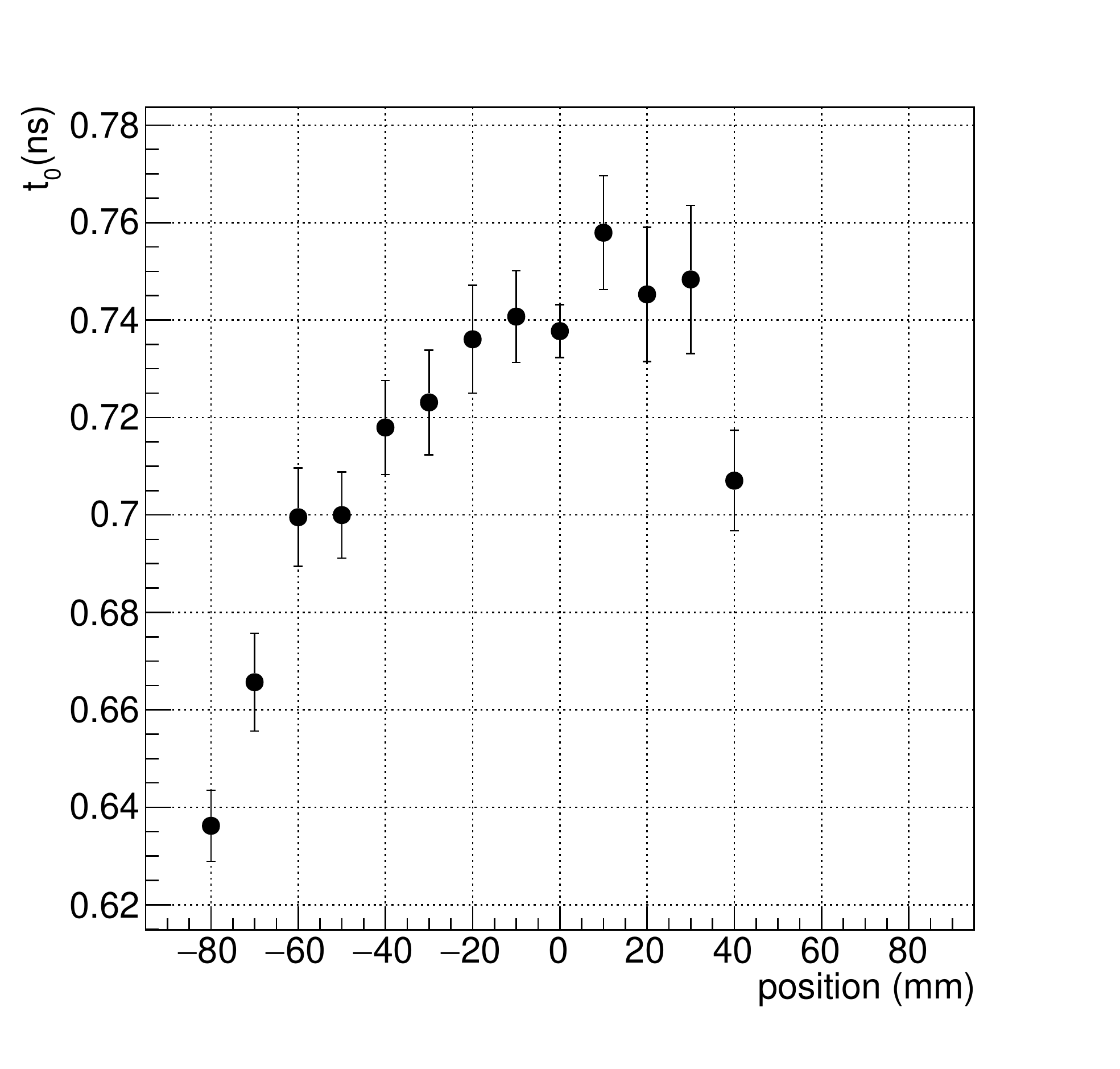}
		\caption{}
		\end{subfigure}
		\begin{subfigure}[h]{0.29\columnwidth} 
		\includegraphics[width=\columnwidth]{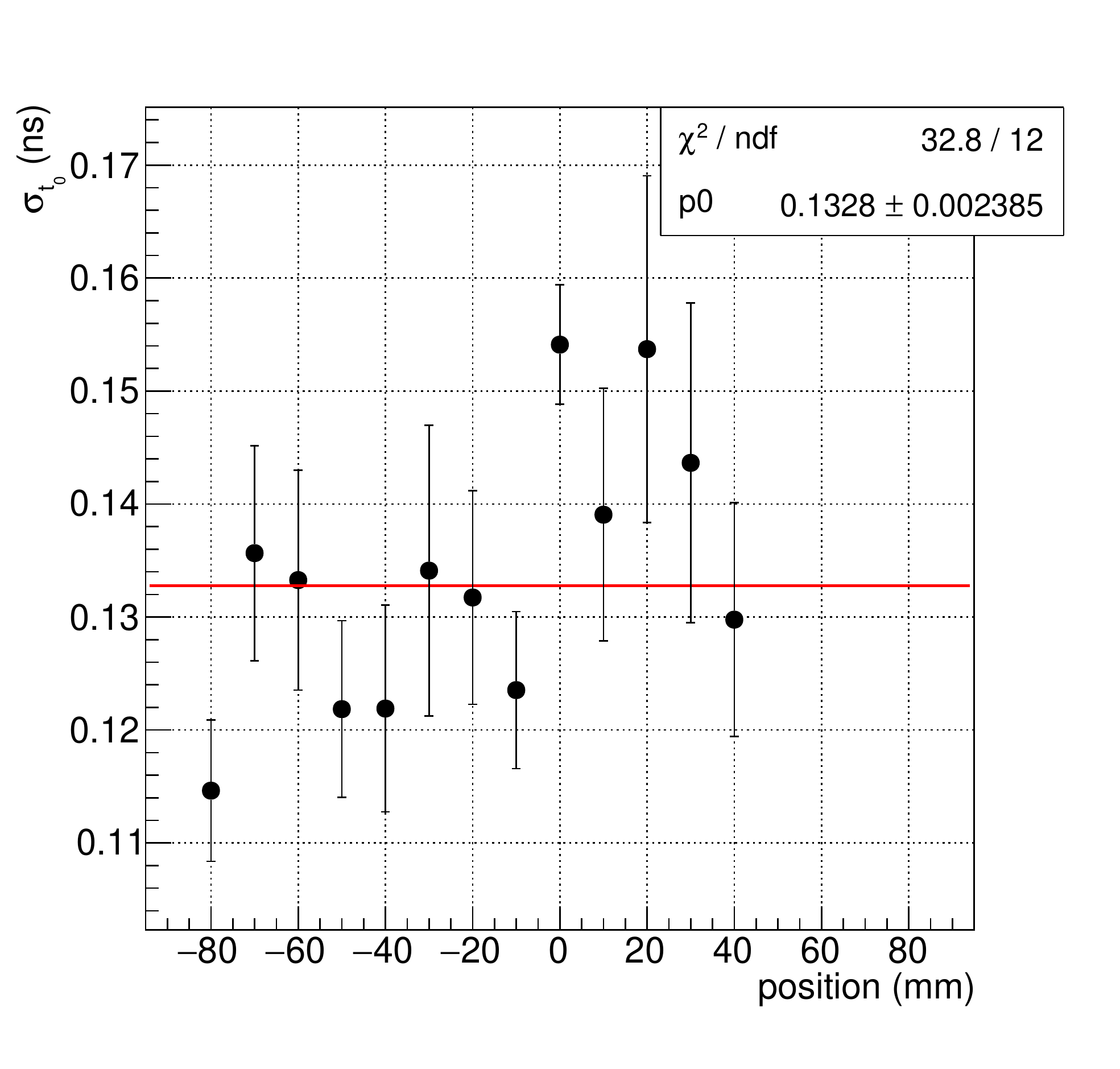}
		\caption{}
		\end{subfigure}
	\caption{For EJ-204 with no reflector:
	(a) The distribution of $t_0$ for one $z$ position.
	(b) The fitted mean $t_0$ vs $z$. (c) The fitted $\sigma_{t_0}$ vs $z$. 
	 }
\label{fig:fig12}
\end{figure}

\begin{figure}[!htbp]
\centering
		\includegraphics[width=0.5\columnwidth]{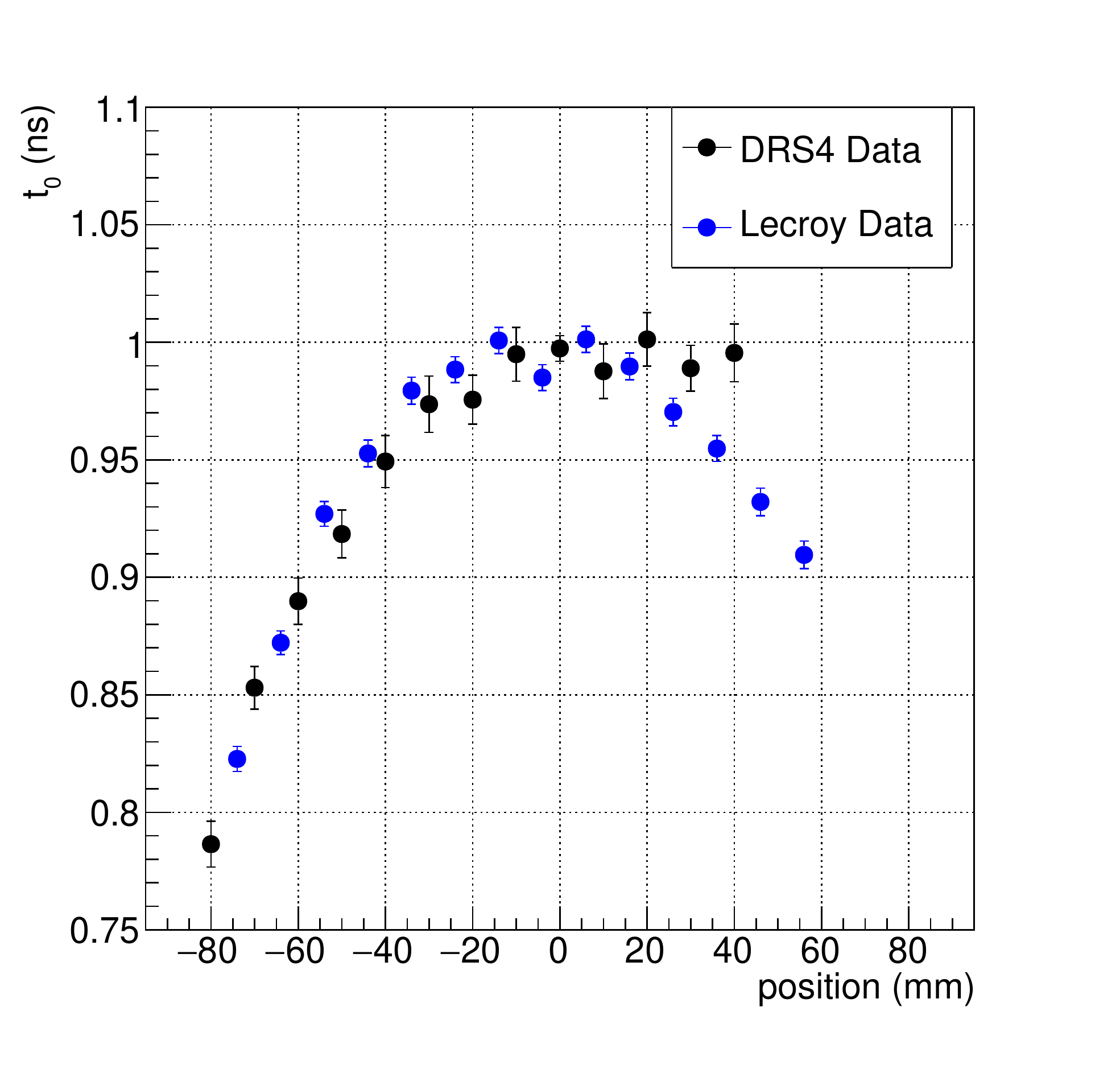}
		\caption{For Teflon-wrapped EJ-204, the interaction time $t_0$ as a function of $z$ position for the DRS4 data acquisition and LeCroy oscilliscope data acquisition. The persistence of the dependence on $z$ indicates that it is not due to potential channel crosstalk in the DRS4 eval board. A constant vertical offset has been applied to the LeCroy data to align it with the DRS4 data, correcting for different timing offsets in the two acquisitions. }
\label{fig:fig13} 
\end{figure}

The overall timing response of the system is evaluated with the tagged $^{22}$Na data, in which the 
interaction time is defined as 
\begin{linenomath*}
\begin{equation}
t_0 =  \frac{t_1+t_2}{2}-t_{tag},
\label{eq:9}
\end{equation}
\end{linenomath*}
where $t_{tag} $ is the pulse time of the tag scintillator, and $t_{1,2}$ are the pulse time of the bar 
interaction evaluated from SiPM1 and SiPM2, respectively. It should be noted that, to increase performance by a factor of
ten compared to traditional two-plane scatter camera systems, we desire an overall timing resolution of roughly 1~ns ($\sigma$)~\cite{svsc}. 
The distribution of $t_0$ for the center of 
the EJ-204, bare measurement is shown in Figure \ref{fig:fig12}a, where only interactions in which the energy deposition is between 300--400~keVee are included. 
The interaction time defined in this manner 
is expected to be constant as a function of the z-position, if we assume a constant velocity throughout the bar, as in Equations \ref{eq:7} and \ref{eq:8}:
\begin{linenomath*}
\begin{eqnarray}
t_0 &=& \frac{z}{2v} + \frac{L-z}{2v} - t_{tag} \nonumber \\
&=& \frac{L}{2v} - t_{tag},
\end{eqnarray}
\end{linenomath*}
However, as Figure \ref{fig:fig12}b shows, this 
is not the case.  Nor is the standard deviation constant, as shown in Figure \ref{fig:fig12}c. This effect, which was seen for all test configurations, may be due 
to the changes in the pulse shape along the rise as the light propagates down the bar. In order to ensure that electronic crosstalk is not
contributing, the test configuration with EJ-204 and Teflon wrapping was repeated with a LeCroy WaveRunner 640zi oscilliscope capturing full waveforms
and applying similar processing methods: 
this interaction time feature persisted, as shown in Figure \ref{fig:fig13}.
While this feature does not directly affect single bar characterizations, it will 
be critical to characterize for the purposes of double scatter imaging. The standard deviation of the 
interaction time $t_0$ is also dependent on the interaction location, however the effect may be statistical due to the 
lower light levels detected for interactions near the center of the bar.  It should be noted that the results are an upper limit on the 
timing resolution, due to the unknown timing resolution of the tag scintillator. While the timing resolution of the tag scintillator 
could be calculated with the assumption of no correlations between $t_{tag}$, $t_{1}$, and $t_{2}$, it is not clear this is a valid assumption. 
The results for all configurations are presented in Table \ref{tab:summary}.

\subsubsection{Timing Response with $^{90}Sr$}

\begin{figure}[!htbp]
\centering		\includegraphics[width=0.9\columnwidth]{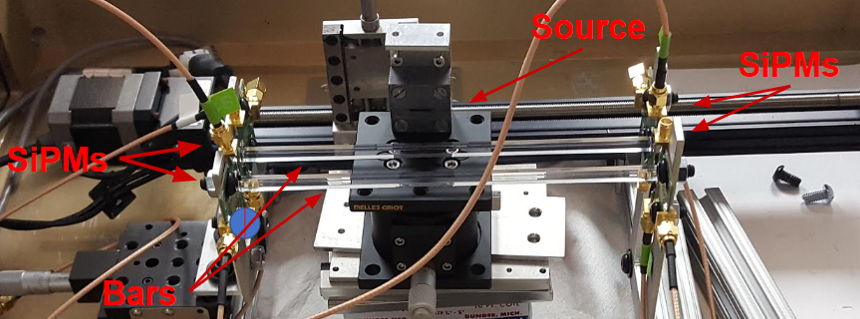}

	\caption{The layout for 2-bar timing tests at UH.}
\label{fig:fig14}
\end{figure}

\begin{figure}[!htbp]
\centering
\begin{subfigure}[h]{0.49\columnwidth}
\includegraphics[width=\columnwidth]{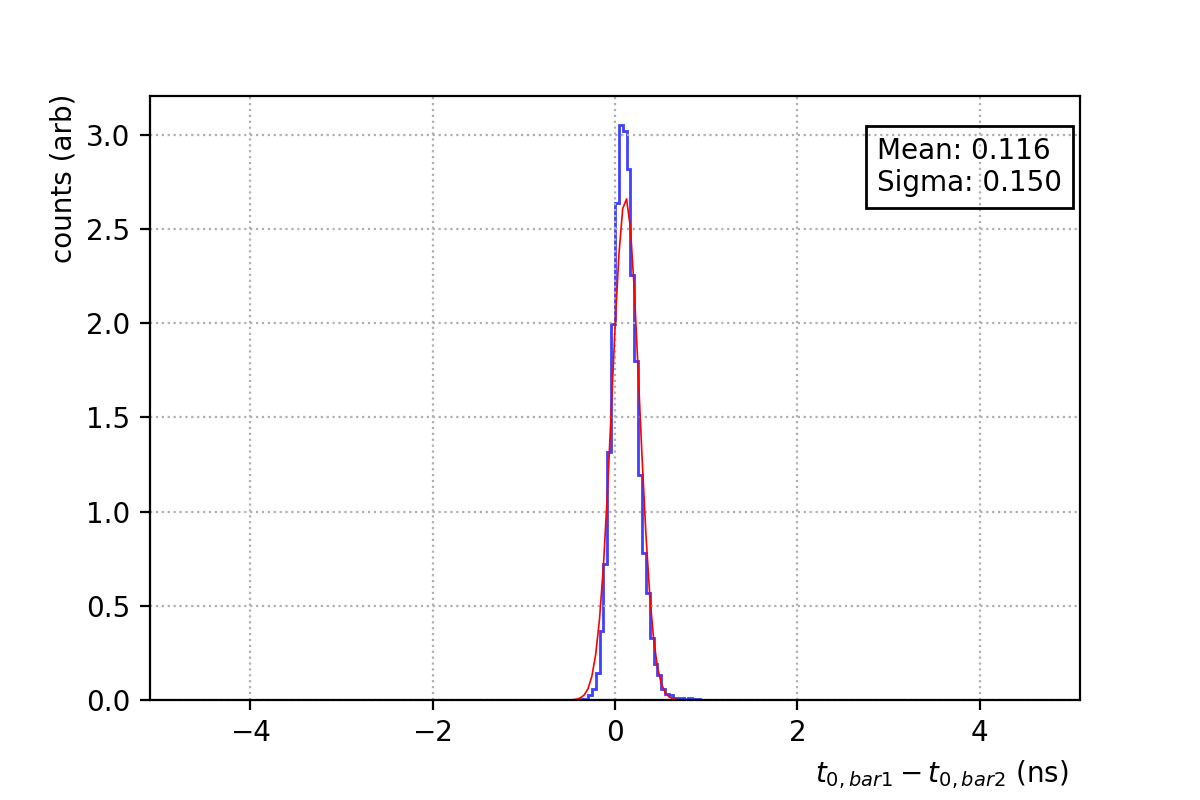}
\caption{}
\end{subfigure}
\begin{subfigure}[h]{0.49\columnwidth}
\includegraphics[width=\columnwidth]{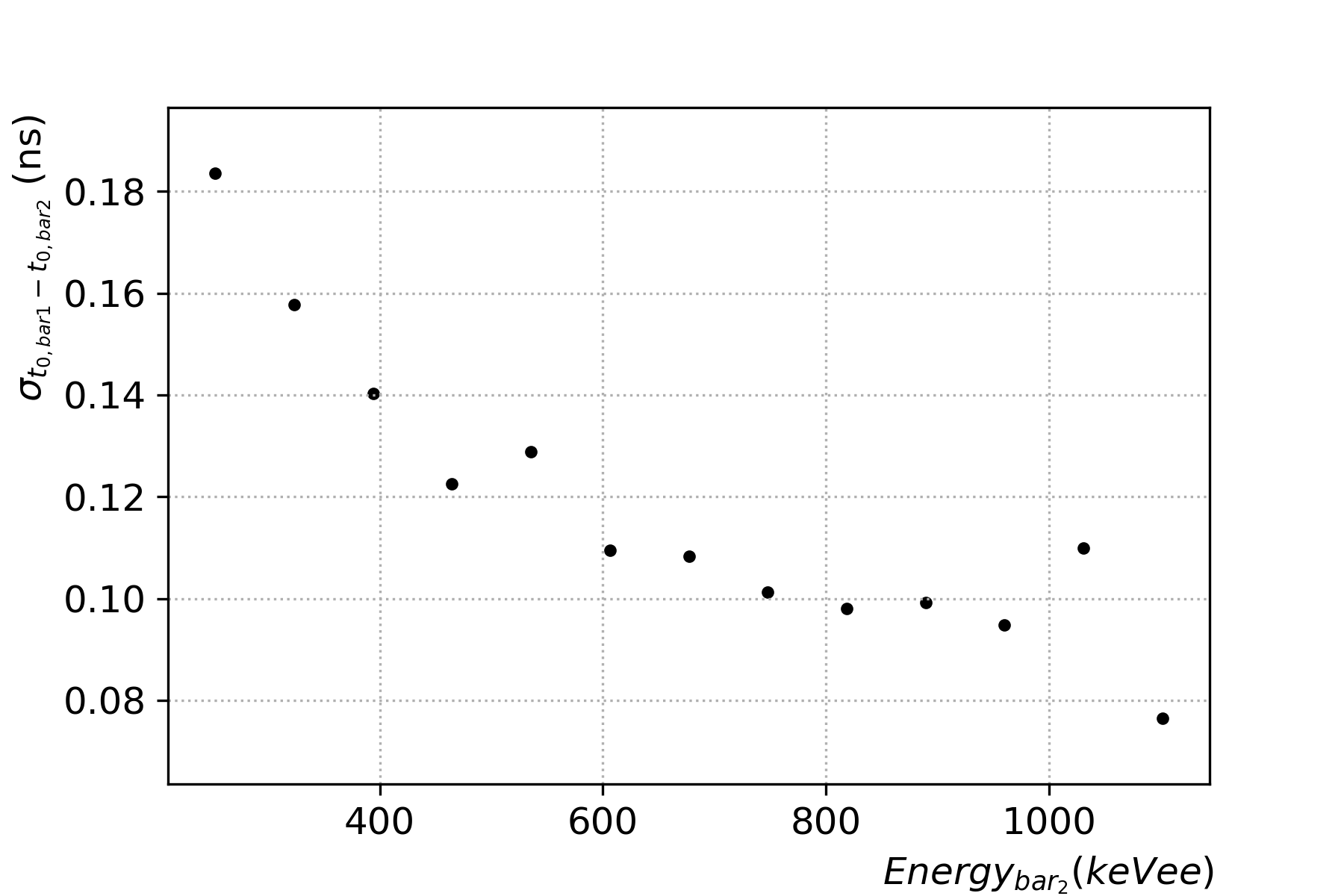}
\caption{}
\end{subfigure}
\caption{ The $t_{0,bar1}-t_{0,bar2}$ distribution, i.e. the arrival time differences in each bar, are shown for a fixed source in (a). This data is then binned versus energy to find a mean and standard deviation, which is plotted as a function of energy deposition in the second bar as shown in (b).}
\label{fig:fig15}
\end{figure}

\begin{figure}[!htbp]
\centering
\begin{subfigure}[h]{0.49\columnwidth}
\includegraphics[width=\columnwidth]{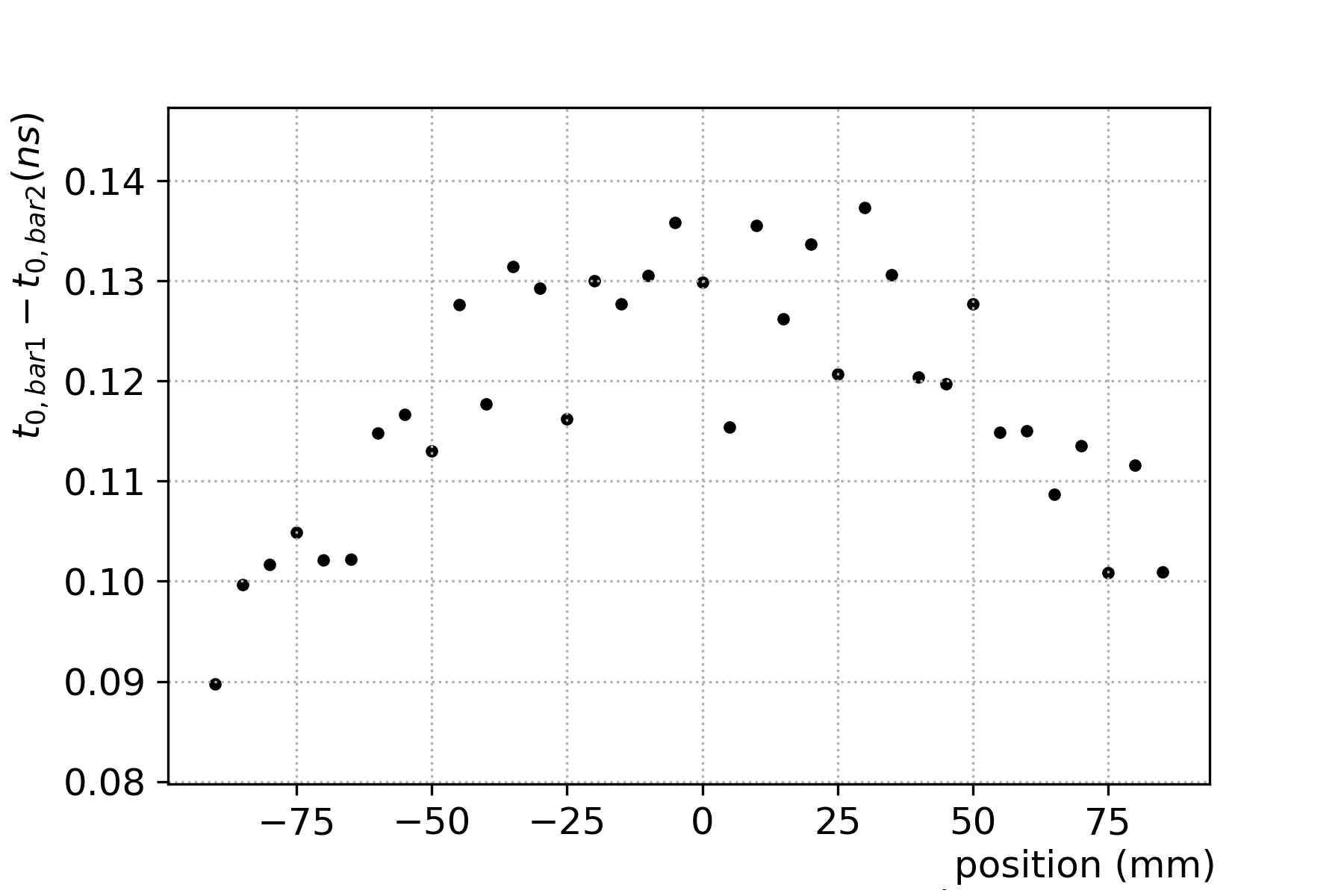}
\caption{}
\end{subfigure}
\begin{subfigure}[h]{0.49\columnwidth}
\includegraphics[width=\columnwidth]{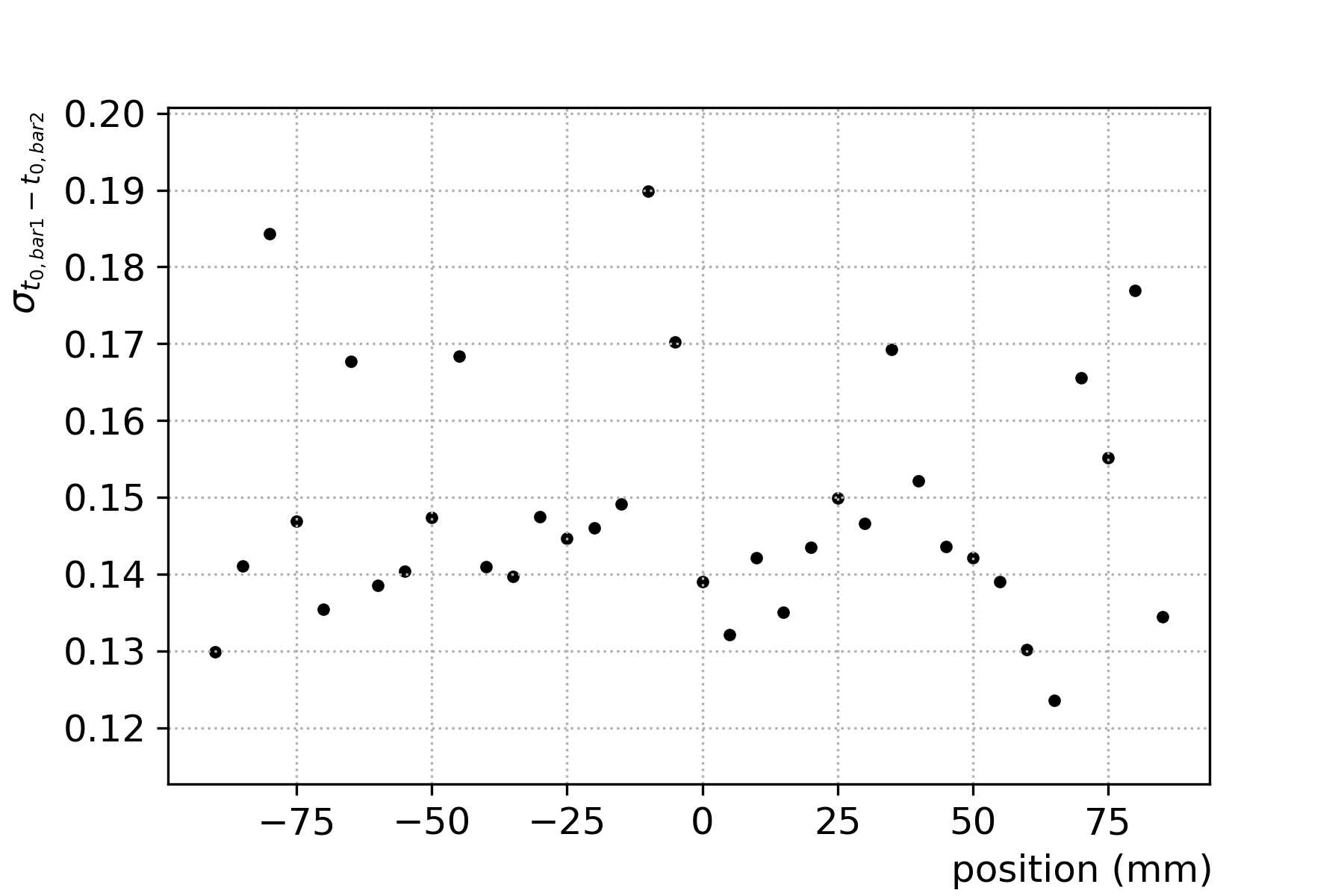}
\caption{}
\end{subfigure}
\caption{The test from Figure \ref{fig:fig15} was repeated with a moving source and the mean and standard deviation of the timing difference as a function of the source position in $z$ are shown in (b) and (c).}
\label{fig:fig16}
\end{figure}

Additional studies were carried out at UH with two scintillator bars triggered in tandem, as shown in Figure \ref{fig:fig14}.
The interaction time was calculated for each bar using Equation \ref{eq:9}, and the difference between the interaction times was studied.
The distribution of interaction time differences, $t_{0,bar1}-t_{0,bar2}$ is shown in Figure \ref{fig:fig15}a for a fixed source position. The timing difference as a function of 
energy deposited in the second bar is shown in Figure \ref{fig:fig15}b.
A further test was carried out with the same setup and a moving $^{90}$Sr source, acquiring 1000 interactions in steps of 50 mm along the length of the bar. 
The mean and standard deviation of the timing distribution as a function of $z$ is shown in Figure \ref{fig:fig16}a and \ref{fig:fig16}b.  Only interactions in which the energy deposition in bar~1 is $1.00\pm0.15$ MeVee are presented, and outliers with timing differences greater than 5 ns are excluded.  

While we observe similar timing resolutions to the $^{22}$Na data, we do not appear to observe the same systematic time shift as a function of interaction position.
We conclude that this is consistent with the hypothesis of pulse shape differences seen in the $^{22}$Na data, as in this case the effect happens on both the front and rear bars, and largely subtracts away.

\section{Summary and Discussions}
Table \ref{tab:summary} summarizes the timing, position, and energy resolution for all configurations tested.
In both amplitude-based and timing-based resolutions, Teflon-wrapped EJ-204 and EJ-230 show the best performance overall.
As the performance is generally in agreement within uncertainties, EJ-204 is identified as the best candidate for neutron kinematic imaging, as it also offers significantly better light yield. 
Because the fission spectrum motivates the lowest possible detection threshold,
we chose EJ-204 over EJ-230 for this particular application. 
The coupling of Teflon and ESR performed similarly in the case of EJ-204, however the difficulties associated with wrapping ESR in a uniform fashion for a 64-bar system, along with optical crosstalk concerns~\cite{esr}, motivated the use of Teflon. 

The results generally suggest that timing-based position calculations should be more robust than amplitude-based calculations due to systematic effects related to optical coupling and triggering. Variations from mechanical construction and optical coupling may have a significant impact on amplitude-based calibrations, and could lead to more complicated requirements on calibration.
We note other effects that also may have significant impact on a full-scale detector designed around such scintillating elements.
For example, the systematic studies indicating variations due to optical coupling, trigger conditions, and structural supports all have significant impact on energy calibrations and light yields.
We have also observed significant differences in performance between repeated measurements of the same scintillator with the same surface treatment, implying that careful calibration of individual elements may be required for a large array of such scintillators. These calibrations will likely need to be done in-situ, since many of our systematic uncertainties are attributed to effects, such as optical coupling, that will arise during assembly of the system. Nevertheless, the data reported here indicate that sub-cm position resolutions can be obtained. 

Finally, we note that although the data presented here show reasonable linearity for the amplitude-based resolutions, prior data on longer bars ($\sim 1$~m length) were better fit by a double exponential. 
This suggests strongly that the optimizations of both scintillator material and wrapping may need to be reevaluated for longer bars, and that the results reported here should not be extrapolated to significantly different geometries due to the complexities of the optical photon propagation.

\begin{table}\begin{center}
\begin{tabular}{|rr||c|c|c|c|c|}
\hline
	\multicolumn{2}{|c||}{\multirow{2}{*}{Scintillator}}	&\multirow{2}{*}{$\sigma_t$ (ps)}&\multicolumn{2}{c|}{$\sigma_z$ (mm)} &\multicolumn{2}{c|}{$\sigma_E/E$ (\%)}\\
\cline{4-7}

	&	&									&$^{22}$Na 			&$^{90}$Sr  	&$^{22}$Na 	&$^{137}$Cs 	 \\	

\hline
\multicolumn{2}{|r||}{EJ-200, bare}	&155$\pm$2 		&13.35			&14.27					&16.7	&14.1\\
				&Teflon		&154$\pm$3		&10.29	 		&7.65					&14.5	&15.8\\
				&ESR		&145$\pm$3		&11.14			&12.09					&16.6	&12.2\\
				\hline
\multicolumn{2}{|r||}{EJ-204, bare}	&136$\pm$3		&10.08			&10.67					&15.7	&14.7\\
				&\bf{Teflon}	&\bf{142$\pm$2}	&\bf{8.06}			&\bf{6.54}					&\bf{13.1}	&\bf{14.3}\\
				&ESR		&125$\pm$3		&8.59			&9.64					&17.6	&12.2\\
				\hline
\multicolumn{2}{|r||}{EJ-230, bare}	&141$\pm$3		&9.61			&8.86					&17.8	&15.0\\
				&Teflon		&142$\pm$2		&8.39			&6.32					&22.6	&13.9\\
				&ESR		&156$\pm$3		&10.17			&8.52					&23.4	&13.0\\
				\hline
\multicolumn{2}{|r||}{EJ-276, bare}&183$\pm$5			&12.13			&13.51					&17.8	&14.1\\
				&Teflon		&171$\pm$2		&9.29			&9.54					&16.5	&14.1\\
				&ESR		&177$\pm$4		&11.65			&10.45					&15.0	&11.3\\
\hline
\multicolumn{2}{|r||}{Syst. error} &$\pm$7		&$\pm$0.73		&$\pm$0.42			&$\pm$3.5			&-\\
\hline
\end{tabular}
\caption{Summary of results with statistical (for $\sigma_t$) and systematic errors for configurations in which they were measured. Note that the energy resolution measurements are at different energies (340 keV for $^{22}$Na and 478 keV for $^{137}$Cs), so differences in resolution are not unexpected. Sufficient data were not acquired for a systematic error of the last column.}
\label{tab:summary}
\end{center}
\end{table}

\section{Conclusions}
We have performed a series of measurements to motivate the choice of scintillator and reflector material for an optically segmented
 scatter-based neutron imaging system. We have determined the top performing scintillator and reflector material in terms of the timing, 
 position, and energy resolution, and overall detectable light to maximize the detection efficiency. In the case of $5\times5\times19~\mathrm{cm}^3$ 
 scintillator bars coupled to SensL's J-series $6\times6~\mathrm{mm}^2$ SiPMs, the top performer was EJ-204 wrapped in Teflon. We expect a 
 threshold of approximately 30~keVee given the light observed and the threshold of the device's electronics.

We have also explored the systematic variability of the position resolution measurement, and have determined that the variability 
of the position as determined by the log ratio of the amplitudes is greater than by the position determined by the difference in 
pulse time. Finally, we observed that the measured interaction time is variable with the interaction location, which is presumed to be biased due to pulse-shape 
distortions on the rising edge as the light propagates along the bar.

\section{Acknowledgments}
Sandia National Laboratories is a multimission laboratory managed and operated by National Technology and Engineering 
Solutions of Sandia, LLC, a wholly owned subsidiary of Honeywell International, Inc., for the U.S. Department of Energy's 
National Nuclear Security Administration under contract DE-NA0003525. This paper describes objective technical results and 
analysis. Any subjective views or opinions that might be expressed 
in the paper do not necessarily represent the views of the U.S. Department of Energy or the United States Government.

We would like to thank the US DOE National Nuclear Security Administration,
Office of Defense Nuclear Non-proliferation for funding this work.

We thank Serge Negrashov for data acquisition software development that was utilized in this work, and Glenn Jocher for prior work on longer scintillating bars that helped to inform this study.

\bibliographystyle{../../../LatexTools/IEEE_style/IEEEtran}

\end{document}